\documentclass[twocolumn,showpacs,preprintnumbers,amsmath,amssymb,superscriptaddress,nofootinbib,english]{revtex4-1}
\usepackage{times,amsmath,amsfonts,amssymb,epstopdf}
\usepackage{graphicx}
\usepackage{dcolumn}
\usepackage{bm}
\usepackage{epsfig}
\usepackage{graphicx}
\usepackage{hyperref}
\usepackage[usenames]{color}
\usepackage{url}
\usepackage[normalem]{ulem}
\usepackage[T1]{fontenc}
\usepackage[dvipsnames]{xcolor}

\newcommand{\tcr}{\textcolor{black}}

\def\be{\begin{equation}}
\def\ee{\end{equation}}
\def\ba{\begin{eqnarray}}
\def\ea{\end{eqnarray}}
\begin{document}

\preprint{IPPP/13/60}
\preprint{DCPT/13/120}

\title{Simulating the quartic Galileon gravity model on adaptively refined meshes}

\author{Baojiu~Li}
\email[Email address: ]{baojiu.li@durham.ac.uk}
\affiliation{Institute for Computational Cosmology, Department of Physics, Durham University, Durham DH1 3LE, UK}

\author{Alexandre~Barreira}
\email[Email address: ]{a.m.r.barreira@durham.ac.uk}
\affiliation{Institute for Computational Cosmology, Department of Physics, Durham University, Durham DH1 3LE, UK}
\affiliation{Institute for Particle Physics Phenomenology, Department of Physics, Durham University, Durham DH1 3LE, UK}

\author{Carlton~M.~Baugh}
\affiliation{Institute for Computational Cosmology, Department of Physics, Durham University, Durham DH1 3LE, UK}

\author{Wojciech~A.~Hellwing}
\affiliation{Institute for Computational Cosmology, Department of Physics, Durham University, Durham DH1 3LE, UK}
\affiliation{Interdisciplinary Centre for Mathematical and Computational Modelling, University of Warsaw, Warsaw, Poland}

\author{Kazuya Koyama}
\affiliation{Institute of Cosmology and Gravitation, University of Portsmouth, Portsmouth PO1 3FX, UK}

\author{Silvia~Pascoli}
\affiliation{Institute for Particle Physics Phenomenology, Department of Physics, Durham University, Durham DH1 3LE, UK}

\author{Gong-Bo~Zhao}
\affiliation{Institute of Cosmology and Gravitation, University of Portsmouth, Portsmouth PO1 3FX, UK}
\affiliation{National Astronomical Observatories, Chinese Academy of Science, Beijing 100012, China}

\date{\today}

\begin{abstract}

We develop a numerical algorithm to solve the high-order nonlinear derivative-coupling equation associated with the quartic Galileon model, and implement it in a modified version of the {\sc ramses} $N$-body code to study the effect of the Galileon field on the large-scale matter clustering. The algorithm is tested for several matter field configurations with different symmetries, and works very well. This enables us to perform the first simulations for a quartic Galileon model which provides a good fit to the cosmic microwave background (CMB) anisotropy, supernovae and baryonic acoustic oscillations (BAO) data. Our result shows that the Vainshtein mechanism in this model is very efficient in suppressing the spatial variations of the scalar field. However, the time variation of the effective Newtonian constant caused by the curvature coupling of the Galileon field cannot be suppressed by the Vainshtein mechanism. This leads to a significant weakening of the strength of gravity in high-density regions at late times, and therefore a weaker matter clustering on small scales. We also find that without the Vainshtein mechanism the model would have behaved in a completely different way, which shows the crucial role played by nonlinearities in modified gravity theories and the importance of performing self-consistent $N$-body simulations for these theories.

\end{abstract}


\maketitle

\section{Introduction}

\label{sect:introduction}

Modified gravity theories \cite{cfps2011} have been widely studied as a possible alternative to the dark energy scenarios \cite{cst2006} (including the standard cosmological constant, or $\Lambda$CDM, paradigm) to explain the observed speed up of the cosmic expansion \cite{gsetal2010,pretal2009,bbetal2011,rsetal2012,hletal2012,rmetal2009}. There has been growing interest in this area recently because study in such models not only sheds light on our understanding of the origin and nature of the cosmic acceleration, but also provides tests of the standard gravity theory, general relativity (GR), on cosmic scales. Such tests are crucial to establish GR as the theory of gravity of the whole universe, valid also on cosmological scales. 

Regardless of passing the cosmological tests, however, any theory of gravity must satisfy a number of stringent local, or solar system, constraints \cite{w2006}, as GR does. Such a requirement is highly nontrivial, because the universality of gravity theory means that if it is modified on cosmic scales, changes can usually also be expected on much smaller scales. Therefore, any viable modified gravity theory must have some dynamical mechanism by which such modifications are suppressed and GR is recovered in high-density (or deep-potential) regions such as the Solar system. Such mechanisms are known as `screening mechanisms' in the literature. The screening mechanism makes gravity behave in different ways in different environments, and this environmental dependence often implies strong nonlinearity in the associated field equations, making studies of these models numerically particularly challenging.

Many of the modified gravity theories studied so far have one or more dynamical scalar-type (spin-0) fields to mediate the modified gravitational force\footnote{See, \cite{lms2008,lms2009,lsb2011a,lsb2011b}, however, for examples with non-dynamical scalar degrees of freedom as origins of modified gravity.}. There are two major classes of such theories. In the first class, the screening is realised by a nonlinear coupling specifying the interaction between the scalar field and matter (or curvature), together with a nonlinear potential specifying the scalar field self-interaction. With appropriate choices of the coupling and potential, the scalar degree of freedom can become very heavy or extremely weakly coupled with matter in dense regions, so that it hardly mediates any interaction between matter particles. The chameleon \cite{kw2004,ms2007} (including $f(R)$ gravity model \cite{sf2010,dt2010b}, see also \cite{lb2007,hs2007,bbds2008}), dilaton \cite{bbds2010} and symmetron \cite{hk2010} models are well-known examples in this class. Out of these models, $f(R)$ gravity (notably the model of \cite{hs2007}) is the most well studied. There have been a number of works which investigated structure formation in this model in the nonlinear regime, with the help of $N$-body simulations \cite{o2008,olh2008,sloh2009,lz2009,zmlhf2010,lz2010,zlk2011,zlk2011b,lh2011,lzk2012,lkzl2012,lzlk2012,jblzk2012,lhkzjb2012,llkz2013,hlcf2013,hlj2013,zwjlw2013,fpsb2013,glm2013}. To facilitate the study, an $N$-body code, {\sc ecosmog} \cite{ecosmog},  was developed, based on the publicly available code {\sc ramses} \cite{ramses}. This code is efficiently parallelised using {\sc mpi} and makes large and high-resolution simulations of $f(R)$ gravity feasible (see \cite{mggadget,isis} for other recent developments of modified gravity simulation codes). Using a parameterisation for this class of modified gravity theories \cite{bdl2011,bdlw2012}, {\sc ecosmog} has also been generalised to study generic chameleon, dilaton and symmetron models \cite{bdlwz2012,bdlwz2013}.

The second class of scalar-type modified gravity theories, with the Dvali-Gabadadze-Porrati (DGP) model \cite{dgp} as a representative example, achieves the screening through nonlinear derivative self-couplings of the scalar field. Here, the nonlinearity causes interference between the gravitational field of individual particles, which makes the deviation from the standard gravity per particle weaker when one keeps adding particles into a system. This is known as the Vainshtein mechanism \cite{vainshtein}, which was first introduced in the context of massive gravity to suppress the extra helicity modes of the massive graviton so as to recover GR in the massless limit. The Vainshtein mechanism works not only in the nonlinear massive gravity \cite{massivegravity1,massivegravity2,cp2012} and DGP models, but also in other more general models, such as the Galileons \cite{nrt2009,dev2009, Galileon1, Galileon2, Galileon3, Galileon4,knt2013}, which are the focus of this paper. The Galileon model is the general name of a class of models whose Lagrangians respect the Galilean shift symmetry, namely the Lagrangian is invariant under the following transformation in the Minkowski spacetime,
\begin{eqnarray}
\partial_\mu\varphi &\rightarrow& \partial_\mu\varphi + b_\mu,
\end{eqnarray}
where $\varphi$ is the Galileon field and $b_\mu$ is a constant four-vector. If one requires that the field equation contains at most second order derivatives, then there are three\footnote{These are in addition to the standard quintessence Lagrangian with a linear potential for the scalar field, which also respects the Galilean shift symmetry (see Eq.~(\ref{eq:5L}) below for more details).} possibilities, respectively the cubic, quartic and quintic Galileons, named after the highest power of $\varphi$ in the Lagrangian (e.g., the quintic Galileon Lagrangian contains the fifth power of $\varphi$, quartic Galileon fourth and cubic Galileon third; more details are given below).

There have been many studies of the astrophysical and cosmological implications of Galileon models (e.g., \cite{ck2009,sk2009,gs2010,dt2010,ndt2010,ags2010,bbd2011,al2012,al2012b,blbp2012,nrcpagb2013,blsbp2013,ott2013,bbbm2013,ls2013}). In particular, there have recently been several $N$-body simulations of these models (e.g., \cite{{schmidt2009,schmidt2009b,cs2009,kw2009},lzk2013,wjl2013,blhbp2013}). However, so far, nonlinear simulation of the Galileon models have been restricted to the DGP and cubic Galileon models, which involve up to second order powers of the scalar field derivatives, e.g., $\left(\nabla^2\varphi\right)^2$ and $\nabla^i\nabla^j\varphi\nabla_i\nabla_j\varphi$, in their field equation. We emphasise that even for these simple Galileon models, those derivative couplings already make the equation highly nonlinear. In the full Galileon model, one can have even higher order derivative coupling terms, such as $\left(\nabla^2\varphi\right)^3$, $\nabla^2\varphi\nabla^i\nabla^j\varphi\nabla_i\nabla_j\varphi$ and $\nabla^i\nabla^j\varphi\nabla_j\nabla_k\varphi\nabla^k\nabla_i\varphi$ in the quartic Galileon model, and we are unaware of any self-consistent simulations of nonlinear structure formation in this case. Recently, Ref.~\cite{blbp2013} has made a spherical collapse study and found that the highest-order Galileon model, the quintic Galileon, admits no physical solution to the Galileon field in the quasi-static limit except when the matter density perturbation is small\footnote{Although this might merely be due to the terms which are neglected in the quasi-static and weak-field approximations used to derive the field equation (as we shall discuss below), it could also imply a breakdown of the model itself. Furthermore, these terms make the full equations of the quartic and quintic Galileon models too complicated and not tractable in practice.}. Therefore, the quintic Galileon model is not of interest to us here, and this work will focus on the $N$-body simulation of the most \tcr{general model that is still viable under the quasi-static approximation, namely} the quartic one. 


In this work we develop an extension of the numerical algorithm used in \cite{lzk2013,blhbp2013} for the DGP and cubic Galileon simulations, and apply it to the quartic Galileon model. This is then implemented in the {\sc ecosmog} code to carry out the first $N$-body simulations of the nonlinear structure formation in this model. Our results show that the nonlinearity due to the high-order derivative couplings plays a very important role in determining the strength of gravity and therefore the clustering of matter, and enable us to quantify the deviations from the standard gravity predictions.

This paper is organised as follows: In \S~\ref{sect:model} we briefly describe the quartic Galileon model and derive its field equations in the quasi-static and weak-field limits; we also introduce the attractor tracker solution of the model, which helps us to remove one Galileon model parameter, obtain analytical expressions for the background quantities and to simplify the field equations. In \S~\ref{sect:equations} we re-express the field equations in the {\sc ramses} code unit and derive the discrete versions of these equations which can be solved on a mesh. We point out that the quartic Galileon equation can be considered as a third-order algebraic equation for $\nabla^2\varphi$, and that first analytically solving the latter could greatly simplify the numerical solutions to the former. A new algorithm is proposed based on this principle, which is shown to work very well for a number of test cases in \S~\ref{sect:code_tests}. Then in \S~\ref{sect:cosmo} we discuss further issues that arise in cosmological simulations, describe the simulations and present our results. We finally summarise and conclude in \S~\ref{sect:summary}.

Throughout the paper we shall follow the metric convention $(+,-,-,-)$, and set $c=1$ except in expressions where $c$ appears explicitly. Greek indices run over $0,1,2,3$ while Roman indices run over $1,2,3$. $M_{\rm Pl}$ is the reduced Planck mass and is related to Newton's constant, $G$, by $M^{-2}_{\rm Pl}=8\pi G$.

\section{The quartic Galileon model}

\label{sect:model}

This section gives a short description of the quartic Galileon model and the derivation of its field equations in appropriate limits under which $N$-body simulations are usually carried out.

\subsection{The model}

The full Galileon model, constrained by the requirement of having at most second-order derivatives in the field equations, is described by the following \tcr{modified Einstein-Hilbet} action
\begin{eqnarray}
S &=& \int{\rm d}^4x\sqrt{-g}\left[\frac{R}{16\pi G}-\frac{1}{2}\sum^5_{i=1}c_i\mathcal{L}_i - \mathcal{L}_m\right],
\end{eqnarray}
in which $R$ is the Ricci scalar, $g$ is the determinant of the metric tensor $g_{\mu\nu}$, $\mathcal{L}_m$ is the Lagrangian density for normal matter fields (photons, neutrinos, baryons and cold dark matter) and $\mathcal{L}_i~(i=1,\cdots,5)$ are the five allowed components of the Galileon Lagrangian density specified by the constant coefficients, $c_i$, which are free parameters of the model. These terms are given by
\begin{eqnarray}\label{eq:5L}
\mathcal{L}_1 &=& M^3\varphi,\nonumber\\
\mathcal{L}_2 &=& \nabla^\mu\varphi\nabla_\mu\varphi,\nonumber\\
\mathcal{L}_3 &=& \frac{2}{M^3}\Box\varphi\nabla^\mu\varphi\nabla_\mu\varphi,\nonumber\\
\mathcal{L}_4 &=& \frac{1}{M^6}\nabla^\lambda\varphi\nabla_\lambda\varphi\Big[2\left(\Box\varphi\right)^2 - 2\nabla^\mu\nabla^\nu\varphi\nabla_\mu\nabla_\nu\varphi\nonumber\\
&& ~~~~~~~~~~~~~~~~~~~~~~~~-\frac{1}{2}R\nabla^\mu\varphi\nabla_\mu\varphi\Big],\nonumber\\
\mathcal{L}_5 &=& \frac{1}{M^9}\nabla^\lambda\varphi\nabla_\lambda\varphi\Big[\left(\Box\varphi\right)^3 - 3\Box\varphi\nabla^\mu\nabla^\nu\varphi\nabla_\mu\nabla_\nu\varphi\nonumber\\
&& ~~~~~~~~~~~~~~~~~~~~~~~~ +2\nabla^\mu\nabla^\nu\varphi\nabla_\mu\nabla_\rho\varphi\nabla^\rho\nabla_\mu\varphi\nonumber\\
&& ~~~~~~~~~~~~~~~~~~~~~~~~ - 6G_{\rho\nu}\nabla^\mu\nabla^\nu\varphi\nabla_\mu\varphi\nabla^\rho\varphi\Big],
\end{eqnarray}
where $G_{\mu\nu}$ is the Einstein tensor, $\varphi$ is the Galileon field and $M$ is a new mass scale characterising the onset of the accelearation epoch, which is defined by $M^3\equiv H_0^2M_{\rm Pl}$, with $H_0$ being the present-day Hubble expansion rate.

Note that $\mathcal{L}_{1,2}$ are the Lagrangian densities for the normal quintessence field with a linear potential. The remaining three terms, $\mathcal{L}_{3-5}$, are characterised by the exponent of $\varphi$ which appears in them, e.g., $\mathcal{L}_5$ is called the quintic Galileon model because it contains the fifth power of $\varphi$. Likewise, $\mathcal{L}_3$ is called the cubic Galileon and $\mathcal{L}_4$ the quartic Galileon. Note, however, that when we talk about the quartic Galileon model we generally set $c_2\neq0, c_3\neq0$, and similarly the quintic model has $c_2,c_3, c_4\neq0$. $c_1$ is always set to $0$ in our study.

Nonlinear structure formation for the cubic Galileons has been systematically studied in \cite{blhbp2013}, which demonstrated that the Vainshtein mechanism is very efficient in suppressing the modified gravity on small scales. As for the quintic Galileons, \cite{blbp2013} showed that the strongly nonlinear equation does not admit real physical solutions in regions where the density contrast is higher than $\mathcal{O}(1)$ (at least in the quasi-static and weak-field limit), and so the model does not merit or support a fully nonlinear study. Consequently, the quartic model is the only cosmologically feasible Galileon model that hasn't been studied with $N$-body simulations, and this will be the main goal of this paper.


The fully covariant expressions of the Galileon energy momentum tensor that the Galileon equation are extremely long and we shall not present them here. Interested readers are referred to, e.g., the appendix of \cite{blbp2012} for the complete formulae. In what follows we shall only give the simplified versions of these equations under appropriate limits.

The background Friedmann and quartic Galileon equations are respectively given by
\begin{eqnarray}\label{eq:friedmann}
3H^2 &=& 8\pi G\bar{\rho}_m + \frac{1}{2}c_2\dot{\varphi}^2 + 6\frac{c_3}{H_0^2}H\dot{\varphi}^3 + \frac{45}{2}\frac{c_4}{H_0^4}H^2\dot{\varphi}^4,\ \  
\end{eqnarray}
and 
\begin{eqnarray}\label{eq:eom}
0 &=& c_2(\ddot{\varphi}+3H\dot{\varphi}) + \frac{c_3}{H_0^2}\left(12H\dot{\varphi}\ddot{\varphi}+6\dot{H}\dot{\varphi}^2+18H^2\dot{\varphi}^2\right)\nonumber\\
&& + \frac{c_4}{H_0^4}\left(54H^2\dot{\varphi}^2\ddot{\varphi}+36\dot{H}H\dot{\varphi}^3+54H^3\dot{\varphi}^3\right),
\end{eqnarray}
in which $\bar{\rho}_m$ is the background density of matter (radiation is neglected because we consider only the late-time universe throughout this paper), $H=\dot{a}/a$ is the Hubble expansion rate and $\varphi$ is the background value of the Galileon field (we do not write overbars here to lighten the notation, see below). A dot is the physical time derivative. Note that from here on we have made a redefinition of the Galileon field
\begin{equation}
\frac{\varphi}{M_{\rm Pl}}\ \rightarrow\ \varphi,
\end{equation} 
so that the new Galileon field $\varphi$ is dimensionless. Correspondingly, $M^3=H_0^2M_{\rm Pl}$ has been used to derive Eqs.~(\ref{eq:friedmann}, \ref{eq:eom}). 

\subsection{Equations in the quasi-static and weak-field limit}

In $N$-body simulations of modified gravity models, we usually work under the quasistatic limit, which means that all time derivatives of the scalar field perturbations are assumed to be small compared with their spatial derivatives ($|\dot{\delta\varphi}|\ll|\delta\varphi_{,i}|$) and can therefore be dropped. In the cases of Galileon models, the quasi-static approximation has been shown to work pretty well on small and intermediate scales in the linear perturbation regime \cite{blbp2012}. For the quartic Galileon, we shall in addition assume that the time derivatives of the gravitational potentials are much smaller than their corresponding spatial derivatives: 
\begin{equation}
|\dot{\Phi}|\sim|\dot{\Psi}|\ll|\Phi_{,i}|\sim|\Psi_{,i}|,~~~\ddot{\Phi}\sim H\dot{\Phi}\ll|\Phi^{,i}_{\ ,i}|,
\end{equation}
where $\Phi, \Psi$ are the Newtonian gauge potentials in the metric
\begin{eqnarray}\label{eq:metric}
{\rm d}^2s &=& (1+2\Psi){\rm d}t^2 - a^2(1-2\Phi)\gamma_{ij}{\rm d}x^i{\rm d}x^j,
\end{eqnarray}
and $_{,i}$ denotes derivative with respect to the comoving coordinate $x^i$. $\gamma_{ij}$ is the metric of the three-dimensional Euclidian space with signature $(+,+,+)$. 

In addition to the quasi-static approximation, we also work in the weak-field limit, which amounts to dropping terms such as $\varphi^{,i}\varphi_{,i}$ compared with $\varphi^{,i}_{\ ,i}$. Note that relaxing the quasi-static and weak-field approximations will result in many new terms entering the final equations, making them much more difficult to derive \tcr{or solve}. We will comment more on the implications of these approximations later.

In the rest of the paper, we use $\nabla_i$ to denote partial spatial derivatives. After some tedious calculation, we find that the (00)-component of the perturbed Einstein equation, when the Galileon contributions are included, can be written as follows in the quasi-static limit (some useful expressions are given in Appendix~\ref{appendix:a}, and note that $\nabla^2\equiv\nabla^i\nabla_i$)
\begin{eqnarray}
\nabla^2\Phi &=& 4\pi G\delta\rho_ma^2 - \frac{c_3}{H_0^2}\dot{\varphi}^2\nabla^2\varphi\nonumber\\
&&+ \frac{3}{2}\frac{c_4}{H_0^4}\frac{1}{a^2}\dot{\varphi}^2\left[\left(\nabla^2\varphi\right)^2-\nabla^i\nabla^j\varphi\nabla_i\nabla_j\varphi\right]\nonumber\\
&&-6\frac{c_4}{H_0^4}H\dot{\varphi}^3\nabla^2\varphi+\frac{3}{2}\frac{c_4}{M^6}\dot{\varphi}^4\nabla^2\Phi.
\end{eqnarray}
Note that to lighten the notation we choose not to use $\delta\varphi$ as the Galileon perturbation; instead, $\varphi$ denotes the full Galileon field in $\nabla\varphi$ and its background value in $\dot{\varphi}$ (the latter is expressed according to the quasi-static approximation, without which the above equation would have been much longer).

At this point, it is useful to introduce some new notation. The quantity $\nabla_i\nabla_j\varphi$ can be decomposed into a trace part and a trace-less part in the following way
\begin{eqnarray}
\nabla_i\nabla_j\varphi &\equiv& \frac{1}{3}\gamma_{ij}\nabla^2\varphi + \bar{\nabla}_i\bar{\nabla}_j\varphi.
\end{eqnarray}
The above equation serves as a definition of the operator $\bar{\nabla}$, which satisfies $\gamma^{ij} \bar{\nabla}_i\bar{\nabla}_j\varphi = 0$. To write down the above decomposition we have used the fact that  ${\nabla}_i{\nabla}_j\varphi$ is symmetric as partial derivatives commute with one another (without such a property we would have to add an anti-symmetric part to the decomposition). An advantage of using this decomposition is that, as we shall see below, when we discretise quantities containing ${\nabla}_i{\nabla}_j\varphi$ on a 3-dimensional mesh, the trace part involves only the central cell and its 6 direct neighbours (i.e., the cells with a common face with the central cell), while the traceless part involves the 6 direct neighbours and the 12 neighbouring cells with a common edge with the central cell. Indeed, we shall see below that this decomposition guarantees that the central cell only enters the Einstein and Galileon equations via the quantity $\nabla^2\varphi$, which is why the operator-splitting technique \cite{cs2009} works for simulations of the DGP model \cite{lzk2013} and the cubic Galileon model \cite{blhbp2013}.

With the above decomposition, we have 
\begin{eqnarray}
\left(\nabla^2\varphi\right)^2-\nabla^i\nabla^j\varphi\nabla_i\nabla_j\varphi &=& \frac{2}{3}\left(\nabla^2\varphi\right)^2-\bar{\nabla}^i\bar{\nabla}^j\varphi\bar{\nabla}_i\bar{\nabla}_j\varphi,\nonumber
\end{eqnarray}
and thus the (00)-component of the Einstein equation can be rewritten as
\begin{eqnarray}\label{eq:00}
\nabla^2\Phi &=& 4\pi G\delta\rho_ma^2 - \frac{c_3}{H_0^2}\dot{\varphi}^2\nabla^2\varphi\nonumber\\
&&+ \frac{c_4}{H_0^4}\frac{1}{a^2}\dot{\varphi}^2\left[\left(\nabla^2\varphi\right)^2-\frac{3}{2}\bar{\nabla}^i\bar{\nabla}^j\varphi\bar{\nabla}_i\bar{\nabla}_j\varphi\right]\nonumber\\
&&-6\frac{c_4}{H_0^4}H\dot{\varphi}^3\nabla^2\varphi+\frac{3}{2}\frac{c_4}{M^6}\dot{\varphi}^4\nabla^2\Phi.
\end{eqnarray}

Similarly, the diagonal part of the ($ij$)-component of the Einstein equation reads
\begin{eqnarray}\label{eq:ij_trace}
&&\nabla^2(\Psi-\Phi)\nonumber\\ 
&=& -\frac{1}{3}\frac{c_4}{H_0^4}\frac{1}{a^2}\dot{\varphi}^2\left[\left(\nabla^2\varphi\right)^2-\frac{3}{2}\bar{\nabla}^i\bar{\nabla}^j\varphi\bar{\nabla}_i\bar{\nabla}_j\varphi\right]\nonumber\\
&&+\frac{c_4}{H_0^4}\left(6\dot{\varphi}^2\ddot{\varphi}+2H\dot{\varphi}^3\right)\nabla^2\varphi\nonumber\\
&&+\frac{3}{2}\frac{c_4}{H_0^4}\dot{\varphi}^4\nabla^2\Psi +  \frac{1}{2}\frac{c_4}{H_0^4}\dot{\varphi}^4\nabla^2\Phi,
\end{eqnarray}
where we have used the fact that dust-like matter and the $c_3$ term Galileon do not contribute to the total pressure perturbation in this limit.

The off-diagonal part of the ($ij$)-component of the Einstein equation is
\begin{eqnarray}\label{eq:ij_traceless}
&&\bar{\nabla}_i\bar{\nabla}_j(\Psi-\Phi)\nonumber\\
&=& \frac{c_4}{H_0^4}\frac{1}{a^2}\dot{\varphi}^2\left[2\bar{\nabla}_i\bar{\nabla}_k\varphi\bar{\nabla}_j\bar{\nabla}^k\varphi - \frac{2}{3}\nabla^2\varphi\bar{\nabla}_i\bar{\nabla}_j\varphi\right]\nonumber\\
&&-\frac{2}{3}\frac{c_4}{H_0^4}\frac{1}{a^2}\dot{\varphi}^2\gamma_{ij}\bar{\nabla}^k\bar{\nabla}^l\varphi\bar{\nabla}_k\bar{\nabla}_l\varphi\nonumber\\
&&+\frac{c_4}{H_0^4}\left(6\dot{\varphi}^2\ddot{\varphi}+2H\dot{\varphi}^3\right)\bar{\nabla}_i\bar{\nabla}_j\varphi\nonumber\\
&&+\frac{3}{2}\frac{c_4}{H_0^4}\dot{\varphi}^4\bar{\nabla}_i\bar{\nabla}_j\Psi+\frac{1}{2}\frac{c_4}{H_0^4}\dot{\varphi}^4\bar{\nabla}_i\bar{\nabla}_j\Phi.
\end{eqnarray}

Finally, the Galileon equation can be written as follows
\begin{widetext}
\begin{eqnarray}\label{eq:eom_nonlinear}
0 &=& -c_2\nabla^2\varphi - \frac{c_3}{H_0^2}\left(4\ddot{\varphi}+8H\dot{\varphi}\right)\nabla^2\varphi + \frac{4}{3}\frac{c_3}{H_0^2}\frac{1}{a^2}\left[\left(\nabla^2\varphi\right)^2-\frac{3}{2}\bar{\nabla}^i\bar{\nabla}^j\varphi\bar{\nabla}_i\bar{\nabla}_j\varphi\right] - 2\frac{c_3}{H_0^2}\dot{\varphi}^2\nabla^2\Psi\nonumber\\
&& + \frac{c_4}{H_0^4}\bigg[\frac{1}{a^4}\bigg(-\frac{4}{9}\big(\nabla^2\varphi\big)^3 + 2\nabla^2\varphi\bar{\nabla}^i\bar{\nabla}^j\varphi\bar{\nabla}_i\bar{\nabla}_j\varphi - 4\bar{\nabla}_i\bar{\nabla}_j\varphi\bar{\nabla}^j\bar{\nabla}^k\varphi\bar{\nabla}_k\bar{\nabla}^i\varphi\bigg)\nonumber\\
&& ~~~~~~~~~~~~+\frac{1}{a^2}(4\ddot{\varphi}+4H\dot{\varphi})\bigg(\big(\nabla^2\varphi\big)^2-\frac{3}{2}\bar{\nabla}^i\bar{\nabla}^j\varphi\bar{\nabla}_i\bar{\nabla}_j\varphi\bigg)
+ 4\frac{1}{a^2}\dot{\varphi}^2\bigg(\nabla^2\varphi\nabla^2\Psi - \frac{3}{2}\bar{\nabla}^i\bar{\nabla}^j\varphi\bar{\nabla}_i\bar{\nabla}_j\Psi\bigg)\nonumber\\
&& ~~~~~~~~~~~~-\frac{4}{3}\frac{1}{a^2}\dot{\varphi}^2\bigg(\nabla^2\varphi\nabla^2\Phi - \frac{3}{2}\bar{\nabla}^i\bar{\nabla}^j\varphi\bar{\nabla}_i\bar{\nabla}_j\Phi\bigg)
- \big(24H\dot{\varphi}\ddot{\varphi}+12\dot{H}\dot{\varphi}^2+26H^2\dot{\varphi}^2\big)\nabla^2\varphi\nonumber\\
&& ~~~~~~~~~~~~-12H\dot{\varphi}^3\nabla^2\Psi + \big(12\dot{\varphi}^2\ddot{\varphi}+4H\dot{\varphi}^3\big)\nabla^2\Phi\bigg],
\end{eqnarray}
\end{widetext}
where we have used
\begin{eqnarray}
&&\nabla_i\nabla_j\varphi\nabla^j\nabla^k\varphi\nabla_k\nabla^i\varphi\nonumber\\ 
&=& \bar{\nabla}_i\bar{\nabla}_j\varphi\bar{\nabla}^j\bar{\nabla}^k\varphi\bar{\nabla}_k\bar{\nabla}^i\varphi + \nabla^2\varphi\bar{\nabla}^i\bar{\nabla}^j\varphi\bar{\nabla}_i\bar{\nabla}_j\varphi\nonumber\\ 
&& + \frac{1}{9}\left(\nabla^2\varphi\right)^3,\nonumber
\end{eqnarray}
and 
\begin{eqnarray}
\nabla^2\varphi\nabla^i\nabla^j\varphi\nabla_i\nabla_j\varphi
&=& \nabla^2\varphi\bar{\nabla}^i\bar{\nabla}^j\varphi\bar{\nabla}_i\bar{\nabla}_j\varphi + \frac{1}{3}\left(\nabla^2\varphi\right)^3.\nonumber
\end{eqnarray}

In principle, the time-dependent coefficients in Eqs.~(\ref{eq:00}, \ref{eq:ij_trace}, \ref{eq:ij_traceless}, \ref{eq:eom_nonlinear}) depend on space as well, but as mentioned above the quasi-static approximation allows us to ignore their spatial dependence. Rigorously speaking, the time dependence in these coefficients is determined by the initial condition of $\dot{\varphi}$, which means that they must be first solved numerically somewhere else and then used when solving the above equations. As we will see in the next subsection, however, the Galileon model has certain properties which allow us to calculate these coefficients analytically.

Before leaving this subsection, let us mention that we would like to eliminate the terms $\nabla^2\Phi, \nabla^2\Psi$, $\bar{\nabla}i\bar{\nabla}_j\Phi$ and $\bar{\nabla}_i\bar{\nabla}_j\Psi$ from Eq.~(\ref{eq:eom_nonlinear}) such that it contains only the matter terms, the Galileon field and its derivatives. However, by using Eqs.~(\ref{eq:00}, \ref{eq:ij_trace}, \ref{eq:ij_traceless}) we can only remove three of them. To \tcr{eliminate} the fourth, we can use the $0i$-component of the Einstein equation, which is
\begin{eqnarray}
G_{0i}\ =\ R_{0i}\ \doteq\ 2\nabla_i\dot{\Phi} + 2H\nabla_i\Psi\ =\ 8\pi GT_{0i},\nonumber
\end{eqnarray}
where $T_{0i}$ has contributions from both matter and the Galileon field. The problem of this approach is that it involves the time derivative of $\nabla\Phi$ which is not negligible ($|(\nabla\Phi)^\cdot| \sim H|\nabla\Psi|$) but which is not straightforward to evaluate in the quasi-static approximation. An alternative approach is to make use of the Euler equation for dark matter which, in the quasi-static limit, reads
\begin{eqnarray}\label{eq:particle_acceleration}
\dot{v}^i + 2Hv^i +  v^k\nabla_kv^i &=& -\frac{1}{a^2}\nabla^i\Psi,
\end{eqnarray}
in which $v^i\equiv{\rm d}x^i/{\rm d}t$ (recall that $x^i$ is the comoving coordinate) is the peculiar velocity of the dark matter fluid. Lowering the indices using $\gamma_{ij}$ and taking the partial derivative with respect to $x^j$, we get
\begin{eqnarray}
&&-\frac{1}{a^2}\bar{\nabla}_i\bar{\nabla}_j\Psi\nonumber\\ 
&=& \nabla_j\dot{v}_i + 2H\nabla_jv_i + \nabla_jv^k\nabla_kv_i + v^k\nabla_j\nabla_kv_i\nonumber\\
&& -\frac{\gamma_{ij}}{3}\left[\nabla\cdot\dot{v} + 2H\nabla\cdot v + \nabla^kv^l\nabla_kv_l + v^k\nabla_k\nabla\cdot v\right],~~~~~~
\end{eqnarray}
in which $\nabla\cdot\mathbf{v}\equiv\nabla^iv_i$. This equation is neater than the $(0i)$-component of the Einstein equation, which also has contributions from the Galileon field. Furthermore, in $N$-body simulations the term $v^k\nabla_kv^i$ in the Euler equation is usually neglected, which gives us
\begin{eqnarray}\label{eq:euler}
&&-\frac{1}{a^2}\bar{\nabla}_i\bar{\nabla}_j\Psi\nonumber\\ 
&\doteq& \nabla_j\dot{v}_i + 2H\nabla_jv_i - \frac{\gamma_{ij}}{3}\left[\nabla\cdot\dot{v} + 2H\nabla\cdot v\right].
\end{eqnarray}
Eq.~(\ref{eq:euler}) allows us to eliminate $\bar{\nabla}_i\bar{\nabla}_j\Psi$ in Eq.~(\ref{eq:eom_nonlinear}) to obtain an equation which contains only the Galileon field and its derivatives. The drawback of this approach, however, is that in simulations it is not always possible to have tracers of the velocity field, especially in low-density regions.

\subsection{Equations on the tracker}

The background Galileon equation, Eq.~(\ref{eq:eom}), admits an attractor tracker solution \cite{dt2010}, which can be expressed as
\begin{eqnarray}\label{eq:tracker}
H\dot{\varphi}\ =\ {\rm const.}\ \equiv\ \xi H^2_0,
\end{eqnarray}
where $\xi$ is a dimensionless constant. This can be checked by substituting Eq.~(\ref{eq:tracker}) into Eq.~(\ref{eq:eom}).

Observational constraints using linear perturbation observables \cite{blsbp2013} show that the best-fitting Galileon model does follow the tracker solution in the recent past. This allows us to calculate the time-dependent coefficients in Eqs.~(\ref{eq:00}, \ref{eq:ij_trace}, \ref{eq:ij_traceless}, \ref{eq:eom_nonlinear}) as follows.

Multiplying both sides of Eq.~(\ref{eq:friedmann}) by $H^2$, using Eq.~(\ref{eq:tracker}) to eliminate $\dot{\varphi}$ and dividing the resulting equation by $H^4_0$, we get
\begin{eqnarray}\label{eq:friedmann2}
E^4 &=& \Omega_ma^{-3}E^2 + \frac{1}{6}c_2\xi^2 + 2c_3\xi^3 + \frac{15}{2}c_4\xi^4,
\end{eqnarray}
in which $E\equiv H/H_0$ and $\Omega_m$ is the fractional energy density for matter today. Setting $a=1$ in the above equation gives
\begin{eqnarray}\label{eq:xi}
\frac{1}{6}c_2\xi^2 + 2c_3\xi^3 + \frac{15}{2}c_4\xi^4 &=& 1-\Omega_m,
\end{eqnarray}
which can be used to determine the value of $\xi$ given $c_2, c_3, c_4$ and $\Omega_m$. Substituting Eq.~(\ref{eq:xi}) back into Eq.~(\ref{eq:friedmann2}), we get
\begin{eqnarray}
E^4 &=& \Omega_ma^{-3}E^2+1-\Omega_m,
\end{eqnarray}
which gives the Hubble expansion rate at $a$ analytically as
\begin{eqnarray}\label{eq:tracker_H}
\left(\frac{H}{H_0}\right)^2 &=& \frac{1}{2}\left[\Omega_ma^{-3}+\sqrt{\Omega_m^2a^{-6}+4(1-\Omega_m)}\right].~~~~~
\end{eqnarray}
Given the solution to $H(a)$, one can then use Eq.~(\ref{eq:tracker}) to find solution to $\dot{\varphi}(a)$ analytically. We will not present this solution explicitly here.

Taking the time derivative of Eq.~(\ref{eq:tracker}), we further obtain
\begin{eqnarray}
\ddot{\varphi}\ = \ -\frac{\dot{H}}{H}\dot{\varphi}\ =\ \frac{1}{2}\xi H_0^2\frac{\varphi''}{\varphi'},
\end{eqnarray}
where $'\equiv{\rm d}/{\rm d}\ln(a)$.

Now, on the tracker, the (00)-component of the Einstein equation becomes 
\begin{eqnarray}\label{eq:00_tracker}
\nabla^2\Phi &=& \alpha_14\pi G\delta\rho_ma^2 + \alpha_2\nabla^2\varphi\nonumber\\
&&+ \frac{\alpha_3}{(H_0a)^2}\left[\left(\nabla^2\varphi\right)^2-\frac{3}{2}\bar{\nabla}^i\bar{\nabla}^j\varphi\bar{\nabla}_i\bar{\nabla}_j\varphi\right],
\end{eqnarray}
where 
\begin{eqnarray}
\alpha_1 &\equiv& \frac{1}{1-\frac{3}{2}c_4\xi^2\varphi'^2},\nonumber\\
\alpha_2 &\equiv& -\frac{c_3\xi\varphi'+6c_4\xi^2\varphi'}{1-\frac{3}{2}c_4\xi^2\varphi'^2},\nonumber\\
\alpha_3 &\equiv& \frac{c_4\xi\varphi'}{1-\frac{3}{2}c_4\xi^2\varphi'^2},
\end{eqnarray}
are time-dependent dimensionless functions. 

The diagonal portion of the ($ij$)-component of the Einstein equation can be used to express $\nabla^2\Psi$ in terms of the Galileon field and matter density perturbation, as
\begin{eqnarray}\label{eq:ij_trace_tracker}
&&\nabla^2\Psi\nonumber\\ 
&=& \alpha_1\alpha_44\pi G\delta\rho_ma^2 + (\alpha_5+\alpha_2\alpha_4)\nabla^2\varphi\\
&&+ \left(\alpha_4-\frac{1}{3}\right)\frac{\alpha_3}{(H_0a)^2}\left[\left(\nabla^2\varphi\right)^2-\frac{3}{2}\bar{\nabla}^i\bar{\nabla}^j\varphi\bar{\nabla}_i\bar{\nabla}_j\varphi\right],\nonumber\ \ 
\end{eqnarray}
where
\begin{eqnarray}
\alpha_4 &\equiv& \frac{1+\frac{1}{2}c_4\xi^2\varphi'^2}{1-\frac{3}{2}c_4\xi^2\varphi'^2},\nonumber\\
\alpha_5 &\equiv& \frac{3c_4\xi^2\varphi''+2c_4\xi^2\varphi'}{1-\frac{3}{2}c_4\xi^2\varphi'^2}
\end{eqnarray}
are time-dependent dimensionless functions. 

The off-diagonal part of the ($ij$)-component of the Einstein equation could be used to express $\bar{\nabla}_i\bar{\nabla}_j\Psi$ in terms of $\bar{\nabla}_i\bar{\nabla}_j\Phi$ and the Galileon field:
\begin{eqnarray}\label{eq:ij_traceless_tracker}
&&\bar{\nabla}_i\bar{\nabla}_j\Phi\nonumber\\ 
&=& \frac{1}{\alpha_4}\bar{\nabla}_i\bar{\nabla}_j\Psi - \frac{\alpha_5}{\alpha_4}\bar{\nabla}_i\bar{\nabla}_j\varphi + \frac{2}{3}\frac{\alpha_3/\alpha_4}{(H_0a)^2}\nabla^2\varphi\bar{\nabla}_i\bar{\nabla}_j\varphi\nonumber\\
&& - \frac{2\alpha_3/\alpha_4}{(H_0a)^2}\left[\bar{\nabla}_i\bar{\nabla}_k\varphi\bar{\nabla}_j\bar{\nabla}^k\varphi - \frac{1}{3}\gamma_{ij}\bar{\nabla}^l\bar{\nabla}^k\varphi\bar{\nabla}_l\bar{\nabla}_k\varphi\right].~~~~~~~~
\end{eqnarray}

With the above expressions, one can eliminate $\nabla^2\Phi, \nabla^2\Psi$ and  $\bar{\nabla}_i\bar{\nabla}_j\Phi$ in the Galileon field equation, while $\bar{\nabla}_i\bar{\nabla}_j\Psi$ remains. As discussed above, there are at least two ways to get rid of this:
\begin{itemize}
\item Using the $(0i)$-component of the Einstein equation and its derivatives, which will however introduce undesired derivative terms of $\Phi$ (and its time derivatives) into the Galileon equation.
\item Using Eq.~(\ref{eq:particle_acceleration}) and its derivatives, which will introduce the velocity field and its time and spatial derivatives into the Galileon equation; alas the velocity field itself is often poorly reconstructed in low-density regions.
\end{itemize}
Given that neither of the above approaches is ideal, we need to think of other alternatives. One option is to replace $\Psi$ with $\Psi_{\rm GR}$, which is the gravitational potential for the same matter distribution but in standard gravity. To see this, we recall that $\Psi_{\rm GR}$ satisfies
\begin{eqnarray}
\nabla^2\Psi_{\rm GR} &=& 4\pi G\delta\rho_ma^2.
\end{eqnarray}
With this expression, Eq.~(\ref{eq:ij_trace_tracker}) can be rewritten as
\begin{eqnarray}
&&\nabla^2\Psi\nonumber\\
&=& \alpha_1\alpha_4\nabla^2\Psi_{\rm GR} + (\alpha_5+\alpha_2\alpha_4)\nabla^2\varphi\nonumber\\
&&+ \frac{3}{2}\frac{\alpha_3}{(H_0a)^2}\left(\alpha_4-\frac{1}{3}\right)\nabla^i\left(\nabla^2\varphi\nabla_i\varphi-\nabla^j\varphi\nabla_i\nabla_j\varphi\right),\nonumber
\end{eqnarray}
where we have used
\begin{eqnarray}
\left(\nabla^2\varphi\right)^2 &=& \nabla_i\left(\nabla^i\varphi\nabla^2\varphi\right) - \nabla^i\varphi\nabla_i\nabla^2\varphi,\nonumber\\
\nabla^i\nabla^j\varphi\nabla_i\nabla_j\varphi &=& \nabla_i\left(\nabla_j\varphi\nabla^i\nabla^j\varphi\right)- \nabla^i\varphi\nabla_i\nabla^2\varphi.\nonumber
\end{eqnarray}
Integrating the above equation once, we get
\begin{eqnarray}\label{eq:force}
&&\nabla_i\Psi\nonumber\\ 
&=& \alpha_1\alpha_4\nabla_i\Psi_{\rm GR} + \left(\alpha_5+\alpha_2\alpha_4\right)\nabla_i\varphi\\
&&+\frac{3}{2}\frac{\alpha_3}{\left(H_0a\right)^2}\left(\alpha_4-\frac{1}{3}\right)\left(\nabla^2\varphi\nabla_i\varphi-\nabla^j\varphi\nabla_i\nabla_j\varphi\right).\nonumber
\end{eqnarray}
Then, $\bar{\nabla}_i\bar{\nabla}_j\Psi$ can be obtained by taking the derivative of this and subtracting the diagonal part, as
\begin{widetext}
\begin{eqnarray}\label{eq:ij_traceless_tracker2}
\bar{\nabla}_i\bar{\nabla}_j\Psi &=& \alpha_1\alpha_4\bar{\nabla}_i\bar{\nabla}_j\Psi_{\rm GR} + \left(\alpha_5+\alpha_2\alpha_4\right)\bar{\nabla}_i\bar{\nabla}_j\varphi\nonumber\\
&& +\frac{3}{4}\frac{\alpha_3}{\left(H_0a\right)^2}\left(\alpha_4-\frac{1}{3}\right)\left[\bar{\nabla}_i\bar{\nabla}_j\left(\varphi\nabla^2\varphi\right)-\varphi\bar{\nabla}_i\bar{\nabla}_j\nabla^2\varphi-\bar{\nabla}_i\bar{\nabla}_j\left(\nabla^k\varphi\nabla_k\varphi\right)+\nabla^2\varphi\bar{\nabla}_i\bar{\nabla}_j\varphi\right].
\end{eqnarray}
\end{widetext}
From a practical point of view, the use of $\Psi_{\rm GR}$ instead of $\Psi$ in the Galileon equation is indeed more convenient, because $\Psi$ and $\varphi$ are interdependent such that the coupled differential equations for them need to be solved simultaneously, which is impossible in $N$-body simulations. On the other hand, $\Psi_{\rm GR}$ does not depend on $\varphi$, so that one can first solve $\Psi_{\rm GR}$ as usual, and then use its value in the Galileon equation to find $\varphi$. The problem with this method is that it involves fourth derivatives of $\varphi$, such as $\bar{\nabla}_i\bar{\nabla}_j\nabla^2\varphi$, and therefore the resulting Galileon field equation can no longer be regarded as an algebraic equation for $\nabla^2\varphi$. Of course, this should not prevent us from solving the equation using a relaxation algorithm, provided that the boundary conditions are set carefully. However, the discretisation of the fourth-derivative terms involves more cells than that of the second-derivative terms, which can make it harder for the relaxation algorithm to converge (we checked this explicitly by implementing Eq.~(\ref{eq:ij_traceless_tracker2}) into our numerical code and confirmed that the convergence property for a few simple tests was worse than that of the method used below). Therefore, we shall not go for this option, though we will revisit this point in the final remarks in \S~\ref{sect:summary}.

%
%

The Galileon field equation, with all terms involving $\nabla^2\Psi$, $\nabla^2\Phi$ and $\bar{\nabla}_i\bar{\nabla}_j\Phi$ eliminated, reads
\begin{widetext}
\begin{eqnarray}\label{eq:eom_tracker}
0 &=& \frac{c^6}{(H_0a)^4}(\nabla^2\varphi)^3+\frac{\beta_1c^4}{(H_0a)^2}(\nabla^2\varphi)^2 + \left[\beta_2+\beta_3\frac{8\pi G\delta\rho_ma^2}{(H_0a)^2}+\frac{\beta_4c^4}{(H_0a)^4}\bar{\nabla}^i\bar{\nabla}^j\varphi\bar{\nabla}_i\bar{\nabla}_j\varphi\right]c^2\nabla^2\varphi\nonumber\\
&& + \frac{\beta_5c^6}{(H_0a)^4}\bar{\nabla}_i\bar{\nabla}_j\varphi\bar{\nabla}^j\bar{\nabla}^k\varphi\bar{\nabla}_k\bar{\nabla}^i\varphi + \frac{\beta_6c^4}{(H_0a)^2}\bar{\nabla}^i\bar{\nabla}^j\varphi\bar{\nabla}_i\bar{\nabla}_j\varphi + \frac{\beta_7c^4}{(H_0a)^2}\bar{\nabla}^i\bar{\nabla}^j\varphi\bar{\nabla}_i\bar{\nabla}_j\Psi + \beta_88\pi G\delta\rho_m a^2, 
\end{eqnarray}
where we have restored the speed of light $c$ and defined the following time-dependent functions
\begin{eqnarray}
\beta_0 &\equiv& -\frac{4}{9}c_4+4\frac{\alpha^2_3}{\alpha_1}\left(\alpha_4-\frac{2}{3}\right),\nonumber\\
\beta_1 &\equiv& \frac{1}{\beta_0}\left[\frac{4}{3}c_3+4c_4\xi+2c_4\xi\frac{\varphi''}{\varphi'}+6\frac{\alpha_3\alpha_5}{\alpha_1}+6\frac{\alpha_2\alpha_3}{\alpha_1}\bigg(\alpha_4-\frac{1}{3}\bigg)\right],\nonumber\\
\beta_2 &\equiv& \frac{1}{\beta_0}\left[-c_2-8c_3\xi-26c_4\xi^2-\left(2c_3\xi+6c_4\xi^2\right)\frac{\varphi''}{\varphi'}+4\frac{\alpha_2\alpha_5}{\alpha_1}+2\frac{\alpha_2^2\alpha_4}{\alpha_1}\right],\nonumber\\
\beta_3 &\equiv& \frac{2}{\beta_0}{\alpha_3}\left(\alpha_4-\frac{1}{3}\right),\nonumber\\
\beta_4 &\equiv& \frac{1}{\beta_0}\left[2c_4+\frac{4}{3}\frac{\alpha_3^2}{\alpha_1\alpha_4}-6\frac{\alpha_3^2}{\alpha_1}\left(\alpha_4-\frac{2}{3}\right)\right],\nonumber\\
\beta_5 &\equiv& -\frac{4}{\beta_0}\left(c_4+\frac{\alpha_3^2}{\alpha_1\alpha_4}\right),\nonumber\\
\beta_6 &\equiv& -\frac{3}{2\beta_0}\left[\frac{4}{3}c_3+4c_4\xi+2c_4\xi\frac{\varphi''}{\varphi'}+\frac{4}{3}\frac{\alpha_3\alpha_5}{\alpha_1\alpha_4}+2\frac{\alpha_3\alpha_5}{\alpha_1}+2\frac{\alpha_2\alpha_3}{\alpha_1}\bigg(\alpha_4-\frac{1}{3}\bigg)\right],\nonumber\\
\beta_7 &\equiv& -\frac{6}{\beta_0}\frac{\alpha_3}{\alpha_1\alpha_4}\left(\alpha_4-\frac{1}{3}\right),\nonumber\\
\beta_8 &\equiv& \frac{1}{\beta_0}\left(\alpha_2\alpha_4+\alpha_5\right).
\end{eqnarray}
\end{widetext}

\begin{figure*}
\includegraphics[width=7.2in,height=3.6in]{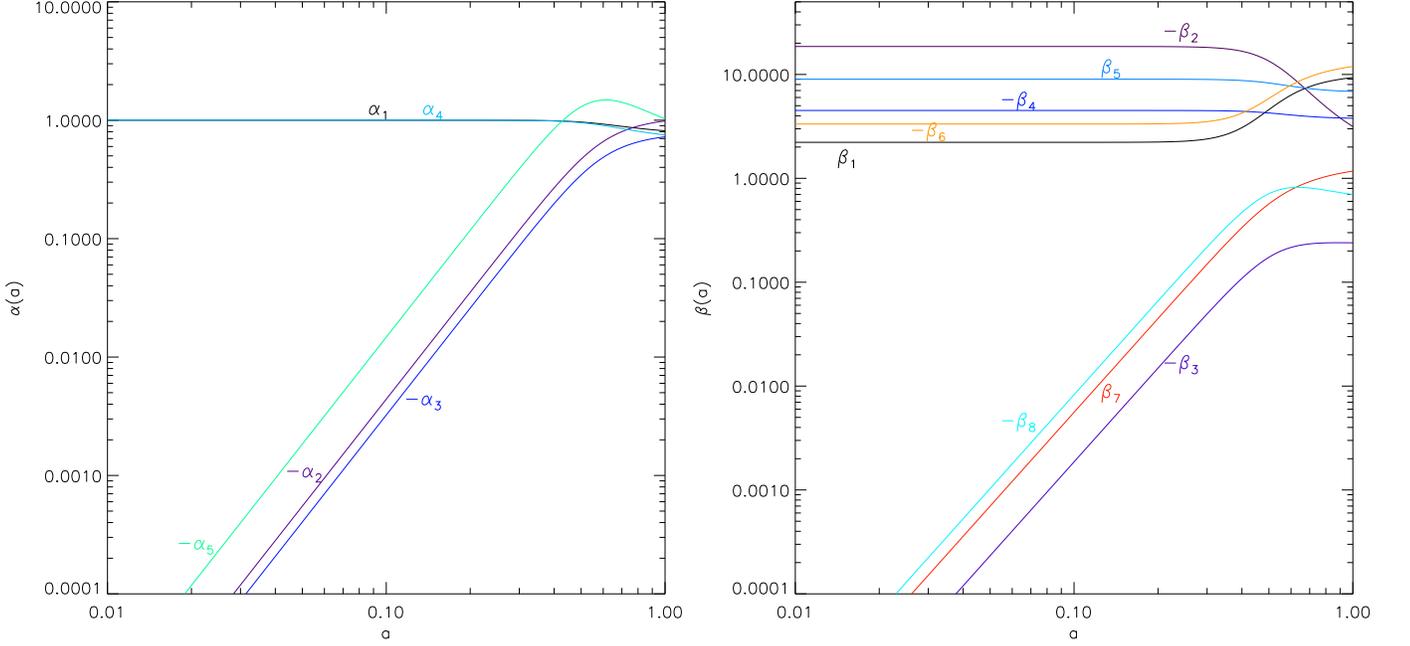}
\caption{(Color online) The time evolution of the $\alpha$ (left) and $\beta$ (right) functions which appear in the field equations, Eqs.~(\ref{eq:ij_traceless_tracker2}, \ref{eq:eom_tracker}), from which we can see that all these quantities are $\mathcal{O}(1-10)$ today.}\label{fig:background}
\end{figure*}

It can be checked that this equation reduces to that of the cubic Galileon model when $c_4=0$. In Eq.~(\ref{eq:eom_tracker}), the term involving $\bar{\nabla}_i\bar{\nabla}_j\Psi$ can be expressed in terms of the velocity field of dark matter, as discussed above, but we shall leave it in its current form for reasons which will become clear below. In Fig.~\ref{fig:background}, we show the time evolution of the $\alpha$- and $\beta$-functions, from which we can see that all of them are of order unity at late times, but $\beta_3$, $\beta_7$ and $\beta_8$ are close to zero at early times.

Eq.~(\ref{eq:eom_tracker}) can be considered as a cubic equation for $\nabla^2\varphi$, of which one can analytically write down the solution. This will be done explicitly in the next section, where we will re-express the equation in the {\sc ramses} code unit.

\section{Discretised equations and algorithm}

\label{sect:equations}

In this section we describe the discretised field equations in appropriate units that the {\sc ecosmog} code solves to carry out cosmological simulations. 

\subsection{The simplified Galileon equation}

The code units used in our code are based on the supercomoving coordinates of \cite{ramses,ms1998}, which can be summarised as follows (tilded quantities are expressed in code unit):
\begin{eqnarray}\label{eq:code_unit}
\tilde{x}\ =\ \frac{x}{B},\ \ \ \tilde{\rho}\ =\ \frac{\rho a^3}{\rho_c\Omega_m},\ \ \ \tilde{v}\ =\ \frac{av}{BH_0},\nonumber\\
\tilde{\Psi}\ =\ \frac{c^2a^2\Psi}{(BH_0)^2},\ \ \ {\rm d}\tilde{t}\ =\ H_0\frac{{\rm d}t}{a^2},\ \ \ \tilde{\varphi}\ =\ \frac{c^2a^2\varphi}{(BH_0)^2},
\end{eqnarray}
in which $x$ is the comoving coordinate, $\rho_c$ is the critical density today, $\Omega_m$ the fractional energy density for matter today and $v$ the particle velocity. In addition, $B$ is the comoving size of the simulation box in unit of $h^{-1}$Mpc. Note that with these conventions the average matter density is $\tilde{\bar{\rho}}=1$. All the above quantities are dimensionless.

In the code unit, Eq.~(\ref{eq:eom_tracker}) can be written as
\begin{eqnarray}\label{eq:eom_code_unit}
0 &=& \left(\nabla^2\varphi\right)^3 + \gamma_1\left(\nabla^2\varphi\right)^2\nonumber\\ 
&& + \left(\gamma_2+\gamma_3\Omega_ma\delta+\gamma_4\bar{\nabla}^i\bar{\nabla}^j\varphi\bar{\nabla}_i\bar{\nabla}_j\varphi\right)\nabla^2\varphi\nonumber\\
&& + \gamma_5\bar{\nabla}_i\bar{\nabla}_j\varphi\bar{\nabla}^j\bar{\nabla}^k\varphi\bar{\nabla}_k\bar{\nabla}^i\varphi + \gamma_6\bar{\nabla}^i\bar{\nabla}^j\varphi\bar{\nabla}_i\bar{\nabla}_j\varphi\nonumber\\
&& + \gamma_7\bar{\nabla}^i\bar{\nabla}^j\varphi\bar{\nabla}_i\bar{\nabla}_j\Psi + \gamma_8\Omega_ma\delta,
\end{eqnarray}
in which we have defined  $\delta \equiv \rho-1$ and the following time-dependent functions
\begin{eqnarray}
\gamma_1 &\equiv& \beta_1a^4,\nonumber\\
\gamma_2 &\equiv& \beta_2a^8,\nonumber\\
\gamma_3 &\equiv& 3\beta_3a^4,\nonumber\\
\gamma_4 &\equiv& \beta_4,\nonumber\\
\gamma_5 &\equiv& \beta_5,\nonumber\\
\gamma_6 &\equiv& \beta_6a^4,\nonumber\\
\gamma_7 &\equiv& \beta_7a^4,\nonumber\\
\gamma_8 &\equiv& 3\beta_8a^8.
\end{eqnarray}
Note that, to lighten the notation, from here on we ignore the tildes and unless otherwise stated all variables are in the code unit.

Being a cubic equation of $\nabla^2\varphi$, Eq.~(\ref{eq:eom_code_unit}) has three branches of solutions, and not all of them are necessarily real. Thus, it is useful to decide which branch to follow prior to solving this equation numerically to find $\varphi$. It is evident that the physical branch of the solutions must reproduce the result $\nabla^2\varphi=0$ in a homogeneous density field, namely when $\delta=0$ in Eq.~(\ref{eq:eom_code_unit}). To write down this solution, let us define
\begin{eqnarray}
\Sigma_1 &\equiv& \gamma_5\bar{\nabla}_i\bar{\nabla}_j\varphi\bar{\nabla}^j\bar{\nabla}^k\varphi\bar{\nabla}_k\bar{\nabla}^i\varphi + \gamma_6\bar{\nabla}^i\bar{\nabla}^j\varphi\bar{\nabla}_i\bar{\nabla}_j\varphi\nonumber\\
&& + \gamma_7\bar{\nabla}^i\bar{\nabla}^j\varphi\bar{\nabla}_i\bar{\nabla}_j\Psi + \gamma_8\Omega_ma\delta,\nonumber\\
\Sigma_2 &\equiv& \gamma_2+\gamma_3\Omega_ma\delta+\gamma_4\bar{\nabla}^i\bar{\nabla}^j\varphi\bar{\nabla}_i\bar{\nabla}_j\varphi,\nonumber\\
\Delta_1 &\equiv& \gamma_1^2-3\Sigma_2,\nonumber\\
\Delta_2 &\equiv& 2\gamma_1^3-9\gamma_1\Sigma_2+27\Sigma_1.
\end{eqnarray}
Then the physical branch of solution is given by (as one could check numerically)
\begin{eqnarray}\label{eq:physical_solution}
\nabla^2\varphi &=& -\frac{1}{3}\left[\gamma_1+2\Delta_1^{1/2}\cos\left(\frac{\Theta}{3}-\frac{2}{3}\pi\right)\right],
\end{eqnarray}
where $\Theta$ is defined by 
\begin{eqnarray}
\cos\Theta &\equiv& \Delta_2/\left(2\Delta_1^{3/2}\right),~~~~\Theta\in[0,\pi)
\end{eqnarray}
This solution is obtained under the condition that
\begin{eqnarray}\label{eq:condition}
4\Delta_1^3-\Delta_2^2 &\geq& 0,
\end{eqnarray}
in which case we find three {\it real} solutions to Eq.~(\ref{eq:eom_code_unit}), but only the above branch gives the correct homogeneous limit. If this condition is violated, there exists only one real solution which does not give the correct physical limit, and this denotes a total breakdown of the model or the approximations which are used to derive our $N$-body equations. 

Note the condition Eq.~(\ref{eq:condition}) automatically guarantees $\Delta_1\geq0$. However, whether Eq.~(\ref{eq:condition}) itself is satisfied or not depends on several factors. There are at least two situations where it is not satisfied (for the best-fit parameters given below):
\begin{itemize}
\item where the density is low (e.g., in voids) -- this is a genuine problem of the quasi-static and weak-field approximations under which our Galileon equation is derived, as demonstrated in a spherical top-hat study of the same model \cite{blbp2013} and in the numerical tests discussed below. This is similar to the problem of imaginary square root in the cubic Galileon model, the cure of which would require us to write down the full Galileon equation dropping all these approximations, which is too complicated and thus beyond the scope of the present work. Here we follow \cite{blhbp2013} and take a simple way out of this problem by setting $4\Delta_1^3-\Delta_2^2 = 0$ whenever Eq.~(\ref{eq:condition}) is violated;
\item where the density field does have physical solutions, but the initial guess of the Galileon field values in the relaxation procedure is bad --  this is a numerical fluke which can be cured, again, by setting $4\Delta_1^3-\Delta_2^2 = 0$ in situations where Eq.~(\ref{eq:condition}) is violated. As the relaxation proceeds, the scalar field moves to its true value  and the problem disappears automatically.
\end{itemize}

The new definitions in Eq.~(\ref{eq:physical_solution}) make the physics rather obscure, but one can still read certain information from it. To see how the screening works, for example, note that when $\delta\gg1$ the terms containing $\delta$ dominate $\Delta_1, \Delta_2$ so that 
\begin{eqnarray}
&&\Delta_1\sim\Delta_2\sim\delta,\nonumber\\
&\Rightarrow& \cos\Theta\sim\delta^{-1/2}\rightarrow0,\nonumber\\
&\Rightarrow& \Theta\rightarrow\frac{1}{2}\pi,\nonumber\\
&\Rightarrow& \cos\left[\frac{1}{3}(\Theta-2\pi)\right]\sim\delta^{-\frac{1}{2}}\nonumber\\
&\Rightarrow& \sqrt{\Delta_1}\cos\left[\frac{1}{3}(\Theta-2\pi)\right] \rightarrow{\rm const.},\nonumber
\end{eqnarray} 
which means that $\nabla^2\varphi$ becomes spatially homogeneous (with a time-varying value). This can also be observed from Eq.~(\ref{eq:eom_code_unit}) with the $\gamma_{3,8}$ terms dominating.


In a similar way, the modified Poisson equation, Eq.~(\ref{eq:ij_trace_tracker}), written in the code unit, becomes
\begin{eqnarray}\label{eq:poisson_code_unit}
&&\nabla^2\Psi\nonumber\\ 
&=& \frac{3}{2}\alpha_1\alpha_4\Omega_ma\delta + (\alpha_5+\alpha_2\alpha_4)\nabla^2\varphi\nonumber\\
&&+\frac{\alpha_3}{a^4}\left(\alpha_4-\frac{1}{3}\right)\left[\left(\nabla^2\varphi\right)^2-\frac{3}{2}\bar{\nabla}^i\bar{\nabla}^j\varphi\bar{\nabla}_i\bar{\nabla}_j\varphi\right].~~~
\end{eqnarray}
Note that, unlike the cases of the cubic Galileon or DGP models, here we have not only the additional terms in the Poisson equation which correspond to the fifth force, but also a time-dependent rescaling of Newton's constant (characterised by the coefficient $\alpha_1\alpha_4$ in front of $\Omega_ma\delta$). This fact has important implications for the behaviour of this model, as we will see below.

\subsection{Discretisation}

We shall use the Gauss-Seidel relaxation algorithm to solve Eq.~(\ref{eq:physical_solution}), which must be discretised beforehand. Let us rewrite this equation in the following form
\begin{eqnarray}
\mathcal{L}\varphi\ \equiv\ \nabla^2\varphi + \frac{1}{3}\left[\gamma_1+2\Delta_1^{1/2}\cos\left(\frac{\Theta}{3}-\frac{2}{3}\pi\right)\right]\ =\ 0, \ \ \ \ \
\end{eqnarray}
where $\mathcal{L}$ is a differential operator (note that in this subsection $\mathcal{L}$ does {\it not} denote the Lagrangian density). 

We will consider a 3-dimensional mesh consisting of cubic cells on which we are to solve the differential equation. Let $h$ denote the side length of each cell and use $\varphi_{i,j,k}$ to denote the value of the Galileon field in the cell which is $i$-th, $j$-th and $k$-th in the $x, y, z$ directions respectively, then the discretisation of some simple field derivatives are given by
\begin{eqnarray}
\nabla_x\varphi &=& \frac{1}{2h}\left(\varphi_{i+1,j,k}-\varphi_{i-1,j,k}\right),\nonumber\\
\nabla_x^2\varphi &=& \frac{1}{h^2}\left(\varphi_{i+1,j,k}+\varphi_{i-1,j,k}-2\varphi_{i,j,k}\right),\nonumber\\
\nabla_x\nabla_y\varphi &=& \frac{1}{4h^2}\big(\varphi_{i+1,j+1,k}+\varphi_{i-1,j-1,k}-\varphi_{i+1,j-1,k}\nonumber\\
 && ~~~~~~~~-\varphi_{i-1,j+1,k}\big).\nonumber
\end{eqnarray}
Starting from these, discretised versions of the more complicated derivative terms which appear in $\Sigma_{1}, \Sigma_2$ can be obtained after a tedious derivation. The expressions are too long for the main text and are put in Appendix~\ref{sect:derivative_discretisation} for interested readers.

\begin{figure*}
\includegraphics[width=7.2in,height=2.4in]{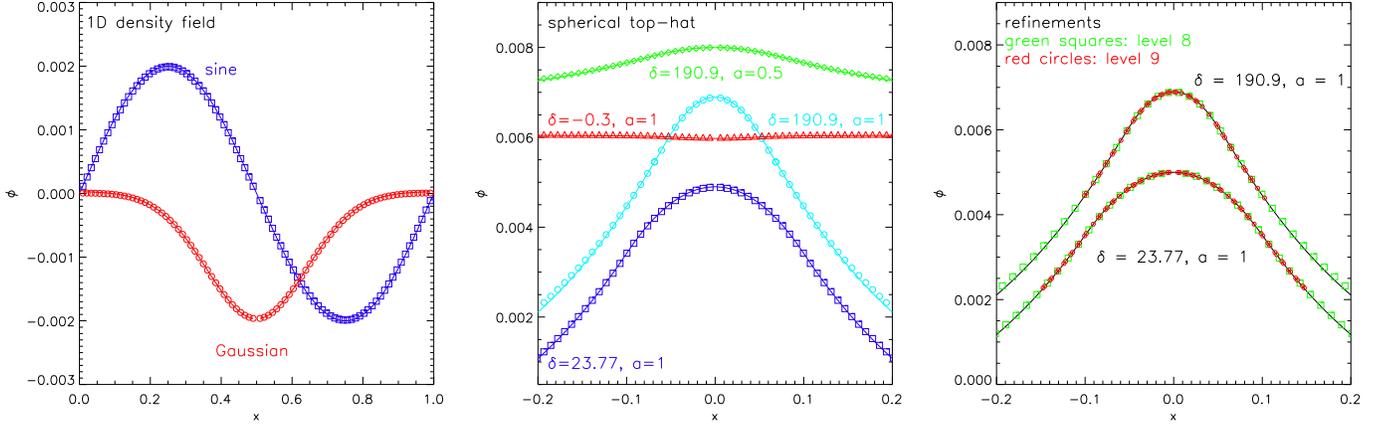}
\caption{(Color online) Tests of the algorithm to solve the Galileon equation. The horizontal axis is the x-coordinate while the vertical axis is the Galileon field value (shifted in the vertical direction arbitrarily to make the  plots clearer). {\it Left}: one-dimensional matter density fields, including a sine-type field, the numerical and analytical (i.e., Eq.~(\ref{eq:1D_sine})) solutions of which are shown as blue squares and blue curve respectively, and a Gaussian-type field, the numerical and analytical (i.e., Eq.~(\ref{eq:1D_gaussian})) solutions are shown as red circles and red curve. {\it Middle}: spherical tophat densities -- the symbols \tcr{mark} our numerical algorithm and the underlying solid curves \tcr{denote} a direct integration of Eq.~(\ref{eq:spherical_analytical}); this test is done for both underdensities and overdensities, at different cosmic times labelled by the expansion factor $a$. {\it Right}: test of refinement levels -- the Galileon equation is solved on a regular level (level 8, green squares) and a refinement level (level 9, red circles) for two tophat overdensities, and the refinements have spherical geometry. For all these tests a regular mesh with $256$ cells in each direction is used. The agreement with the analytical solutions is excellent, showing that our algorithm works very well.}\label{fig:tests}
\end{figure*}

The Galileon differential operator can then be written as
\begin{widetext}
\begin{eqnarray}\label{eq:discrete_op}
\mathcal{L}^h\varphi_{i,j,k} &=& \frac{1}{h^2}\left(\varphi_{i+1,j,k}+\varphi_{i-1,j,k}+\varphi_{i,j+1,k}+\varphi_{i,j-1,k}+\varphi_{i,j,k+1}+\varphi_{i,j,k-1}-6\varphi_{i,j,k}\right)\nonumber\\
&& + \frac{1}{3}\left[\gamma_1+2\Delta_{1,i,j,k}^{1/2}\cos\left(\frac{\Theta_{i,j,k}}{3}-\frac{2}{3}\pi\right)\right],
\end{eqnarray}
\end{widetext}
in which all the derivative-coupling terms are incorporated in $\Sigma_{1,2,i,j,k}$ (through $\Theta_{i,j,k}$). As discussed in Appendix~\ref{sect:derivative_discretisation}, these terms do not involve $\varphi_{i,j,k}$. Therefore, as far as the discrete equation is concerned, the differential operator $\mathcal{L}^h$ is linear so that linear Gauss-Seidel relaxation, or fast Fourier transform, can be used to solve the discretised Galileon equation. This is the advantage of splitting $\nabla_i\nabla_j\varphi$ into a trace and a trace-less part, and it is essentially how the operator-splitting trick \cite{cs2009} works for the DGP model. However, this does not imply that the convergence property for the quartic Galileon equation is necessarily as good as that of the standard Poisson equation, because the quantities $\Sigma_{1,2,i,j,k}$ change during each relaxation sweep, and are not fixed (this is in contrast to the source term of the Poisson equation).

The Gauss-Seidel relaxation updates $\varphi_{i,j,k}$ according to
\begin{eqnarray}
\varphi^{h,{\rm new}}_{i,j,k} &=& \varphi^{h,{\rm old}}_{i,j,k} - \frac{\mathcal{L}^h\left(\varphi^{h,{\rm old}}_{i,j,k}\right)}{\frac{\partial\mathcal{L}^h(\varphi^{h,{\rm old}}_{i,j,k})}{\partial\varphi^{h,{\rm old}}_{i,j,k}}},
\end{eqnarray}
where $\varphi^{h,{\rm old}}_{i,j,k}$ and $\varphi^{h,{\rm new}}_{i,j,k}$ are, respectively, the values of $\varphi_{i,j,k}$ before and after a relaxation sweep, and
\begin{eqnarray}
\frac{\partial\mathcal{L}^h(\varphi^{h,{\rm old}}_{i,j,k})}{\partial\varphi^{h,{\rm old}}_{i,j,k}} &=& -\frac{6}{h^2}.
\end{eqnarray}

The discretisation of the modified Poisson equation is similar, and we shall not present the details here.

\section{Code tests}

\label{sect:code_tests}

In this section we carry out several tests to confirm that the above algorithm implemented in the {\sc ecosmog} code indeed works properly. Such tests are important for us to be confident about the reliability and accuracy of our numerical solutions.

Following \cite{ecosmog}, the simplest nontrivial test one can do to the code is to check that in a homogeneous density field the solution is constant everywhere. In such a test, the initial guess of the scalar field values on the simulation mesh is set randomly. It is then desired that after a few Gauss-Seidel relaxations the field approaches the same value across the whole mesh. We have checked that our code passes this test, but we shall not show the results here. Instead, in the rest of this section we will focus on other more sophisticated tests.

In all tests, we use the best-fit values of $\Omega_m$ and the Galileon parameters ($c_2, c_3, c_4, \xi$) as given below. Further technical details can be found in the caption of Fig.~\ref{fig:tests}.

\subsection{One dimensional density fields}

\label{subsect:1D_dens}

As in the case of DGP and cubic Galileon models, it can be shown that if the density field is one dimensional (e.g., $\delta$ is \tcr{allowed to vary only in the $x$ and not the $y, z$-directions}), then all nonlinear terms in the Galileon field equation vanish and we are left with the following linearised equation
\begin{eqnarray}\label{eq:1D_eqn}
\frac{{\rm d}^2}{{\rm d}x^2}\varphi(x) &=& -\frac{\gamma_8}{\gamma_2}\Omega_ma\delta
\end{eqnarray}
(note that throughout this section all quantities are expressed in the code units). This equation is straightforward to solve in simple density configurations, enabling one to find analytical solutions. Here, we consider two representative examples.

In the first case, we make the ansatz 
\begin{eqnarray}\label{eq:1D_sine}
\varphi(x) &=& A\sin(2\pi x),
\end{eqnarray}
in which $x\in[0,1]$ and $A$ is a constant, use Eq.~(\ref{eq:1D_eqn}) to obtain the corresponding $\delta(x)$, which we then plug into the code and check the solution to $\varphi(x)$ agrees with the original ansatz. The density field is given by
\begin{eqnarray}
\delta(x) &=& \frac{\gamma_2}{\gamma_8}\frac{4\pi^2A}{\Omega_ma}\sin(2\pi x),
\end{eqnarray}
and in the test we take $A=0.002$ for illustration purposes (we have checked other values of $A$ and found similar results).

In the second case, the anstaz is that $\varphi(x)$ has a Gaussian form,
\begin{eqnarray}\label{eq:1D_gaussian}
\varphi(x) &=& A\left\{1-\alpha\exp\left[-\frac{(x-0.5)^2}{w^2}\right]\right\},
\end{eqnarray}
for which the density field is given by
\begin{eqnarray}
\delta(x) &=& \frac{2A\alpha}{w^2}\left[1-2\frac{(x-0.5)^2}{w^2}\right]\exp\left[-\frac{(x-0.5)^2}{w^2}\right].\ \ \
\end{eqnarray}
We take $A=0.002, \alpha=0.9999$ and $w=0.2$.

The left panel of Fig.~\ref{fig:tests} shows the results of these 1D tests, in which the blue squares and red circles are, respectively, the numerical solutions to $\varphi(x)$ in the sine and Gaussian tests. The analytical solutions, Eq.~(\ref{eq:1D_sine}) and Eq.~(\ref{eq:1D_gaussian}), are shown as solid curves with the same colours. One can see that the numerical and analytical solutions are in excellent agreement with each other.

\subsection{Spherical overdensity}

\label{subsect:spherical_dens}

In the 1D tests above, all nonlinear terms do not contribute to the Galileon equation, and to check the behaviour of these terms it is necessary to do tests in which these terms do contribute. The simplest possibility which satisfies this criterion is a spherically symmetric density configuration, such as a top-hat overdensity or underdensity.

With a lengthy but trivial derivation, it can be shown that in the spherically symmetric case the Galileon field equation can be reduced to
\begin{eqnarray}\label{eq:spherical_eqn}
0 &=& \left(\frac{1}{r}\frac{{\rm d}\varphi}{{\rm d}r}\right)^3 + \eta_1\left(\frac{1}{r}\frac{{\rm d}\varphi}{{\rm d}r}\right)^2\nonumber\\ 
&&+ \left[\eta_2+\eta_3\Omega_ma^{-3}\hat{\delta}\right]\left(\frac{1}{r}\frac{{\rm d}\varphi}{{\rm d}r}\right)+\eta_4\Omega_ma\hat{\delta},
\end{eqnarray}
in which $\eta_{1-4}$ are time-dependent functions defined by
\begin{eqnarray}
\eta_0 &\equiv& \frac{1}{a^8}\left[-4c_4+36\frac{\alpha_3^2}{\alpha_1}\left(\alpha_4-\frac{2}{3}\right)\right],\nonumber\\
\eta_1 &\equiv& \frac{1}{\eta_0a^4}\left[4c_3+12c_4\xi+6c_4\xi\frac{\varphi''}{\varphi'}\right]\nonumber\\
&& + \frac{1}{\eta_0a^4}\left[18\frac{\alpha_3\alpha_5}{\alpha_1}+18\frac{\alpha_2\alpha_3}{\alpha_1}\left(\alpha_4-\frac{1}{3}\right)\right],\nonumber\\
\eta_2 &\equiv& -\frac{1}{\eta_0}\left[c_2+8c_3\xi+26c_4\xi^2+\left(2c_3\xi+6c_4\xi^2\right)\frac{\varphi''}{\varphi'}\right]\nonumber\\
&& + \frac{2}{\eta_0}\frac{\alpha_2}{\alpha_1}\left(2\alpha_5+\alpha_2\alpha_4\right),\nonumber\\
\eta_3 &\equiv& \frac{6}{\eta_0}\alpha_3\left(\alpha_4-\frac{1}{3}\right),\nonumber\\
\eta_4 &\equiv& \frac{1}{\eta_0}\left(\alpha_5+\alpha_2\alpha_4\right),
\end{eqnarray}
and $\hat{\delta}$ is the average density {\it within} (not at) radius $r$.

Eq.~(\ref{eq:spherical_eqn}) is a third-order algebraic equation in $\frac{1}{r}\frac{{\rm d}\varphi}{{\rm d}r}$ and, of its three solutions, the physical one (i.e., the one which satisfies $\frac{1}{r}\frac{{\rm d}\varphi}{{\rm d}r}\rightarrow0$ as $\hat{\delta}\rightarrow0$) is given by (see also \cite{blbp2013})
\begin{eqnarray}\label{eq:spherical_analytical}
\frac{1}{r}\frac{{\rm d}\varphi}{{\rm d}r} &=& -\frac{1}{3}\left[\eta_1+2\sqrt{\hat{\Delta}_1}\cos\left(\frac{\hat{\Theta}-2\pi}{3}\right)\right],
\end{eqnarray}
in which we have defined 
\begin{eqnarray}
\cos\hat{\Theta} &\equiv& \frac{\hat{\Delta}_2}{2\sqrt{\hat{\Delta}_1^3}},\nonumber\\
\hat{\Delta}_1 &\equiv& \eta_1^2-3\left(\eta_2+\eta_3\Omega_ma^{-3}\hat{\delta}\right),\nonumber\\
\hat{\Delta}_2 &\equiv& 2\eta_1^3-9\eta_1\left(\eta_2+\eta_3\Omega_ma^{-3}\hat{\delta}\right)+27\eta_4\Omega_ma\hat{\delta}.\ \ \ 
\end{eqnarray}
Eq.~(\ref{eq:spherical_analytical}) could then be easily integrated to obtain $\varphi(r)$. Particularly, $\hat{\delta}$, and therefore $\frac{1}{r}\frac{{\rm d}\varphi}{{\rm d}r}$, is a constant inside a spherical top-hat density configuration, making it easy to find $\varphi$.

In our numerical implementation using a mesh with $256^3$ cubic cells, we put the spherical top-hat of radius $R$ at the centre of the simulation box. The density value inside and outside the top-hat is set in different ways: if the top-hat is overdense, we set $\delta=-0.1$ (note that $\delta$ and $\hat{\delta}$ are different quantities!) in all cells outside $R$ and the value inside $R$ is chosen such that the values of $\delta$ in all cells add to zero; if the top-hat is underdense, we set $\delta=-0.3$ in all cells inside $R$ and the value outside $R$ is chosen such that the values of $\delta$ in all cells, again, add to zero. Correspondingly, we have done tests for 4 different top-hats:
\begin{itemize}
\item $\delta_{\rm in}\approx23.77$, $\delta_{\rm out}=-0.10$, $R=0.10$ and $a=1.0$, where $\delta_{\rm in,out}$ are, respectively, the values of $\delta$ inside and outside the top-hat;
\item $\delta_{\rm in}\approx190.9$, $\delta_{\rm out}=-0.10$, $R=0.05$ and $a=1.0$;
\item $\delta_{\rm in}\approx190.9$, $\delta_{\rm out}=-0.10$, $R=0.05$ and $a=0.5$;
\item $\delta_{\rm in}=-0.30$, $\delta_{\rm out}\approx0.000157$, $R=0.10$ and $a=1.0$.
\end{itemize}
The numerical solutions to these cases are plotted in the middle panel of Fig.~\ref{fig:tests}, respectively as blue squares, cyan circles, green diamonds and red triangles. The corresponding semi-analytical solutions, obtained by integrating Eq.~(\ref{eq:spherical_analytical}), are shown as solid curves with the same colours. Again, the agreement is rather good, except at large radii where the effects of the finite simulation box and periodic boundary conditions becomes non-negligible.

These tests already reveal some physical behaviour of the Galileon field. For example, 
\begin{itemize}
\item given the same overdensity, the Galileon field profile is much deeper at late times than at early times;
\item at a given time, increasing the overdensity value makes the Galileon field profile steeper;
\item the slopes of the Galileon field have opposite signs in overdense and in underdense regions.
\end{itemize}
More importantly, we find that the numerical code fails to converge for very low-density tophats (e.g., $\delta_{\rm in}<-0.5$). This is because in such cases $\cos\hat{\Theta}>1$, which has no solutions at all: the physical branch of solutions to Eq.~(\ref{eq:spherical_eqn}) becomes complex, marking the breakdown of the existence of solutions to the Galileon field equation (there is still one real solution to Eq.~(\ref{eq:spherical_eqn}) but unfortunately this is not the physical solution and therefore cannot be taken). This problem is very similar to the imaginary-value problem found in cubic Galileon simulations, which also appeared in low-density regions \cite{blhbp2013}. As we mentioned above, the origin of this problem probably lies in the quasi-static and weak-field approximations which have been used to derive the $N$-body equations: although the terms we dropped are negligible in high-density regions, they could be important in low-density regions where other terms are small too (and may also cancel one another to some extent) \footnote{Or this can be an inherent problem of the quartic Galileon model, and even if all terms are included there are still portions of the parameter space which have no physical solutions. In general, nonlinear differential equations do not always have solutions and even if they have, the solutions may not be unique.}. 

The derivation of the full equation with all terms included, however, is too \tcr{complicated} for the quartic Galileon model. Therefore, despite the above problem, we shall keep working in the quasi-static and weak-field limits. Instead, we shall introduce some simple fixes to the problem in the cosmological simulations below.

\subsection{Refinements}

\label{subsect:refinements}

Finally, as our code employs adaptive-mesh refinement \cite{ramses} to achieve high resolutions in high-density regions, it is important to check that it works properly on the refinements as well. For this test, we again use spherical top-hats as the underlying density configurations. But instead of having a regular mesh with $256^3$ cells, we refine a region which fully covers the top-hat (by `refine' we mean split a cubic cell into 8 equal-sized cubic sub-cells) and also solve the Galileon field equation on the refinement. We call the two levels `level 8' and `level 9' respectively, because they have $2^8$ and $2^9$ cells in one dimension. On these two levels, the density field is set up in exactly the same way. The boundary conditions on level 9 is set up by fixing the values of the Galileon field in all boundary cells.

In the right panel of Fig.~\ref{fig:tests}, we compare the solutions on the two levels for two density configurations: $\delta_{\rm in}\approx23.77, R=0.10$ (lower) and $\delta_{\rm in}\approx190.9, R=0.05$ (upper), both at $a=1.0$. Numerical solutions on levels 8 and 9 are represented by green squares and red circles respectively, while the black solid curves are the solutions by directly integrating Eq.~(\ref{eq:spherical_analytical}) with respect to $r$. In both cases, we find very good agreement between the solutions on different levels, which shows that the code works well on refinements.

\section{Cosmological simulations}

\label{sect:cosmo}

Next let us turn to cosmological simulations of the quartic Galileon model. Because the field equations for this model are considerably more complicated than any of other modified gravity models which have been studied with simulations so far, several new issues arise here which we shall discuss first. We then describe the simulations and discuss the physical results.

\subsection{Implementations in the cosmological setup}

\label{sect:cosmo_algorithm}

\subsubsection{Calculation of the modified gravity force}

The ultimate purpose of solving the Galileon equation in $N$-body simulations is to compute the modified gravity force and use it to update particle positions and velocities. According to Eq.~(\ref{eq:particle_acceleration}), the total gravitational force is given by $\nabla_i\Psi$ (and not $\nabla_i\Phi$!).

In Newtonian gravity and certain modified gravity theories (such as the DGP and cubic Galileons),  the two gravitational potentials $\Psi$ and $\Phi$ are identical due to the lack of anisotropic stress, and therefore one usually needs to solve the (modified) Poisson equation for $\Phi$. In the quartic Galileon model, $\Phi\neq\Psi$ and so it is more straightforward to solve Eq.~(\ref{eq:ij_trace_tracker}) to obtain $\Psi$ than to solve the modified Poisson equation Eq.~(\ref{eq:00_tracker}) for $\Phi$.

Because of the appearance of derivative self couplings of the scalar field, Eq.~(\ref{eq:ij_trace_tracker}), or its code-unit version, Eq.~(\ref{eq:poisson_code_unit}), is more complicated than the Poisson equation in other theories. There are two methods to deal with these derivative-coupling terms:
\begin{itemize}
\item use Eq.~(\ref{eq:force}) to compute the total gravity force once the Galileon field has been obtained -- this approach is very straightforward in terms of the force calculation, but it does not give $\Psi$, which is needed to solve the Galileon equation, because Eq.~(\ref{eq:force}) cannot be further integrated.
\item compute all the new terms in Eq.~(\ref{eq:poisson_code_unit}), which we define as $\rho_{\rm eff}$, and add $\rho_{\rm eff}$ to the source term of the standard Poisson equation in the $N$-body solver to obtain $\Psi$ {\it and} the total gravity. More explicitly, $\rho_{\rm eff}$ is defined (in code units) as
\begin{eqnarray}\label{eq:rho_eff}
&&\rho_{\rm eff}\nonumber\\ 
&\equiv& \frac{3}{2}\left(\alpha_1\alpha_4-1\right)\Omega_ma\delta + (\alpha_5+\alpha_2\alpha_4)\nabla^2\varphi\nonumber\\
&&+\frac{\alpha_3}{a^4}\left(\alpha_4-\frac{1}{3}\right)\left[\left(\nabla^2\varphi\right)^2-\frac{3}{2}\bar{\nabla}^i\bar{\nabla}^j\varphi\bar{\nabla}_i\bar{\nabla}_j\varphi\right].~~
\end{eqnarray}
\end{itemize}
In our simulations, we follow the second approach.

\subsubsection{$\Psi$ in the Galileon equation}

Until now, we have not been able to eliminate the derivative terms of $\Psi$ from the Galileon equation. As mentioned earlier, there are at least three ways to do this:
\begin{itemize}
\item using the $(0i)$-component of the Einstein equation and its derivative, which will, however, introduce undesired derivative terms of $\Phi$ (and its time derivatives) into the Galileon equation;
\item using Eq.~(\ref{eq:particle_acceleration}) and its derivatives, which will introduce the velocity field and its time and spatial derivatives into the Galileon equation; also the velocity field is usually poorly reconstructed in low-density regions;
\item using Eq.~(\ref{eq:ij_traceless_tracker2}) to eliminate $\bar{\nabla}_i\bar{\nabla}_j\Psi$, which makes the Galileon equation higher order and therefore more difficult to converge in the numerical solutions.
\end{itemize}
As none of the above approaches is ideal, we shall follow a more straightforward alternative, namely using the values of $\Psi$ directly in the Galileon equation. Because the Galileon field is solved before $\Psi$ in the simulation, this means that we have to utilise $\Psi$ from the previous time step to solve the Galileon equation in the current step. This approximation, however, is not a disadvantage of using this method, because
\begin{itemize}
\item even if we use the first two methods above, we still need to know either $\Phi$ or $v$ at the current time step (which we do not know at the time of solving the Galileon equation) to evaluate $\dot{\Phi}$ and $\dot{v}$, and therefore some approximations are needed anyway;
\item the values of $\Psi$ at the previous time step are only needed for the regular simulation mesh, and on refinements we can instead use interpolated values from coarser levels, because the Galileon equation on refinements are solved after $\Psi$ is obtained on the coarser level. 
\end{itemize}
Indeed, we shall discuss later, the error caused by this approximation is small.

In our simulations, we store the values of $\Psi$ for the current time step to be used for the next time step.

\subsubsection{Convergence criterion}

The standard Newton-Gauss-Seidel relaxation method has been used in all versions of {\sc ecosmog}, and there is no need to repeat its introduction here. Interested readers are referred to \cite{ecosmog} for more detail. Instead, here we shall only briefly describe the new features of the new code, the most important of which is the convergence criterion.

In simulations of $f(R)$ gravity, the chameleon, dilaton and symmetron models, and the DGP and cubic Galileon models, we have seen very good convergence properties of the scalar field solver: as more relaxation sweeps are done, the residual (which is the modulus of the difference of the two sides of the scalar field equation) could become very small ($\lesssim10^{-15}$). In the quartic Galileon case, due to the higher degree of nonlinearity (e.g., terms such as $\bar{\nabla}_i\bar{\nabla}_j\varphi\bar{\nabla}^j\bar{\nabla}^k\varphi\bar{\nabla}_k\bar{\nabla}^i\varphi$), we cannot decrease the residual by as much, and below some value the residual starts to oscillate. The problem is not alleviated if we change the colouring scheme (i.e., the order for updating the scalar field in different cells), or use other algorithms such as successive over-relaxation.

Because the residual at a given refinement level is the root-mean-squared of the residuals for all its cells, this implies that in some cells the scalar field value is not as accurate as in other cells. We have checked explicitly and found that this happens mostly when some of the terms in the Galileon equation (i.e., some of the terms specified by $\gamma_1-\gamma_8$) are much larger than other terms, and this appears only for a small fraction of all the cells\footnote{We note that in the simple tests above the residual can often become much smaller. This indicates that the fact that the residual in the simulations cannot get arbitrarily small could be because of the complicated nonlinear density field in real simulations (and the approximation used in dealing with $\bar{\nabla}_i\bar{\nabla}_j\Psi$).}.

Fortunately, one does not have to rely on the residual being extremely small to decide on the convergence of the relaxation method. When the residual is smaller than the truncation error, which is essentially the numerical discretisation error at a given refinement level, further reducing the residual does not really help because it cannot beat the error caused by solving a {\it discrete} version of a {\it continuous} equation \cite{ptvf}. In our cases, when the residual stops decreasing and begins oscillating, it is already at least one or two orders of magnitude smaller than the truncation error -- this more than satisfies the convergence criterion advocated by \cite{ptvf}, namely the residual being smaller than about one third of the truncation error. Therefore, in the simulations, we stop the relaxation in practice when the residual starts to oscillate. The fact that we do not require the residual to become very small makes the overall performance of the Galileon simulations better since, for example, in $f(R)$ simulations a lot of time is spent on reducing the residual to below at least $10^{-12}\sim10^{-10}$.

In what follows, we will also find that the spatial variation of the Galileon field is strongly suppressed in high-density regions, so that its effect on structure formation in these regions is minimal anyway.

\subsubsection{Time integration}

\tcr{The update of particle positions and velocities is done using a second-order midpoint time integration 
scheme as is implemented in the default {\sc ramses} code \cite{ramses}. This reduces to standard
second-order leapfrog scheme but allows the extra freedom of varying lengths of time steps. The lengths 
of time steps are determined by the Courant-Friedrich-Levy stability condition.}

\tcr{Since the inclusion of the fifth force in the quartic Galileon model amounts to a change of the Newtonian 
potential in standard $\Lambda$CDM simulations, the above time-integration scheme also works for our 
simulations provided that the proper total gravitational potential is used. Note that, in cases where gravity 
is stronger than that in GR, the time steps are made shorter to satisfy the stability condition: this ensures
that the accuracy of the time integration is similar to that in {\sc ramses} $\Lambda$CDM simulations.}

\subsection{The cosmological models}

\begin{figure*}
\includegraphics[width=7.2in,height=3.6in]{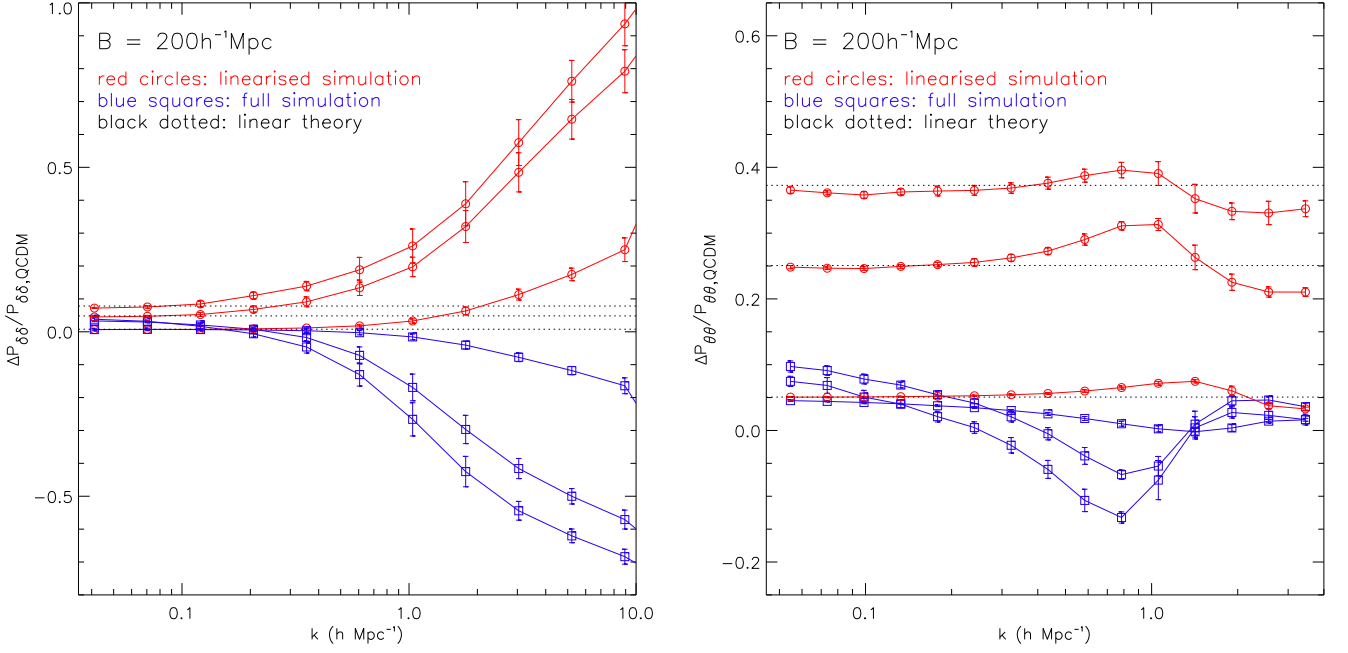}
\caption{(Colour online) The nonlinear matter (left) and velocity divergence (right) power spectra for the models studied in this paper, with box sizes $B=200h^{-1}$Mpc. The horizontal axis is the scale $k$ and the vertical axis is the relative difference of $P(k)$ for the full simulations (blue squares and blue curves) and the linearised simulations (red circles and red curves) from that of the QCDM simulation. The linear theory predictions for $P_{\delta\delta}(k)$ and $P_{\theta\theta}(k)$ are plotted as black dotted horizontal lines in each panel. The results are plotted at three different times: $a=1.0$, $a=0.8$ and $a=0.5$, in order of decreasing deviation from QCDM at small scales (large $k$) for $P_{\delta\delta}$ and at large scales (small $k$) for $P_{\theta\theta}$. All simulation data are averaged over the five realisations, and the error bars show the 1-$\sigma$ standard deviation.}\label{fig:pk}
\end{figure*}

Following most previous Vainshtein simulations,  we shall perform simulations for three different models in this work:
\begin{itemize}
\item the {\it full quartic Galileon} model;
\item the {\it linearised Galileon} model, in which all nonlinear terms of the Galileon equation and the modified Poisson equation are suppressed, such that there is no Vainshtein screening; here the clustering of matter is governed by a scale-independent effective Newton constant given by
\begin{eqnarray}\label{eq:Geff}
\frac{G_{\rm eff}}{G} &=& \alpha_1\alpha_4 - \frac{2\gamma_8}{3\gamma_2}\left(\alpha_5+\alpha_2\alpha_4\right); 
\end{eqnarray}
\item the so-called {\it QCDM} model, which has the same background expansion history as the full Galileon model but in which gravity is not modified.
\end{itemize}
By simulating the linearised Galileon model, one can understand the effect of the Vainshtein mechanism, and by working with the QCDM model one picks out the effects of having the Galileon background evolution history only.

The full Galileon model we consider is the same as the one studied in \cite{blbp2013}, which has been shown to fit the CMB temperature power spectrum data from WMAP9 \cite{hletal2012}, supernovae data from SNLS \cite{gsetal2010} and BAO data from SDSS DR7 \cite{rsetal2012}, 6dF \cite{bbetal2011} and BOSS \cite{pretal2009} very well. Its cosmological parameters are
\begin{eqnarray}
&&\left\{\Omega_ch^2, \Omega_bh^2, h, \tau, n_s, \log\left[10^{10}A_s\right]\right\}\nonumber\\
&=& \left\{0.126, 0.02182, 0.7334, 0.0791, 0.945, 3.152\right\},
\end{eqnarray}
where $h=H_0/(100{\rm km/s/Mpc})$, $\Omega_{b,c}$ are the fractional energy densities for baryons and cold dark matter respectively, $\tau$ is the reionisation optical depth, $n_s, A_s$ the slope and amplitude (at the pivot scale $k=0.02$Mpc$^{-1}$) of the primordial power spectrum; the Galileon model parameters are
\begin{eqnarray}
\left\{c_2/c_3^{2/3}, c_3, c_4/c_3^{4/3}\right\} &=& \left\{-4.55, 20.0, -0.096\right\}.
\end{eqnarray}
For these parameters, our modified {\sc camb} code \cite{camb,blbp2012} calculates that the age of the Universe is 13.77 Gyr and $\sigma_8=0.998$. Although in general the initial condition of the Galileon field, $\dot{{\varphi}}_i$, is also a free parameter, the fact that we follow the tracker solution allows us to fix its value -- using the above parameters of the quartic Galileon model we found that the tracker solution is characterised by $\xi\doteq0.4133$. Note however that in this model the matter density is higher than the best-fit $\Lambda$CDM model for the same data sets, so that even the QCDM model (i.e., without any modification of gravity) gives stronger matter clustering than the latter.

To understand the effect of modified gravity better, all models use the same initial condition, which is generated using the {\sc mpgrafic} \cite{mpgrafic} code at an initial redshift $z_i=49$. We can do this because the modified gravity effect is negligible at $z_i$ so that these models have the same behaviour\footnote{This means that before $z_i$ the growth of density perturbations is governed by $\Lambda$CDM (or rather CDM because the effect of the Galileon field is negligible). However, the matter density and other cosmological parameters are given by the best-fit Galileon model, which are different from those of the WMAP9 best-fit $\Lambda$CDM model.}. We generated five realisations of initial conditions with different phases, each of which with $256^3$ particles.

The simulations are started on a regular 3D mesh with $256$ cubic cells in each direction, and the density field on the mesh is constructed using the triangular-shaped cloud scheme. Cells are refined when the effective number of particles inside them exceeds $8.0$. We have two different box sizes of the simulation: $B=200h^{-1}$Mpc and $B=400h^{-1}$Mpc respectively.

\subsection{First numerical results}

\begin{figure*}
\includegraphics[width=7.2in,height=3.6in]{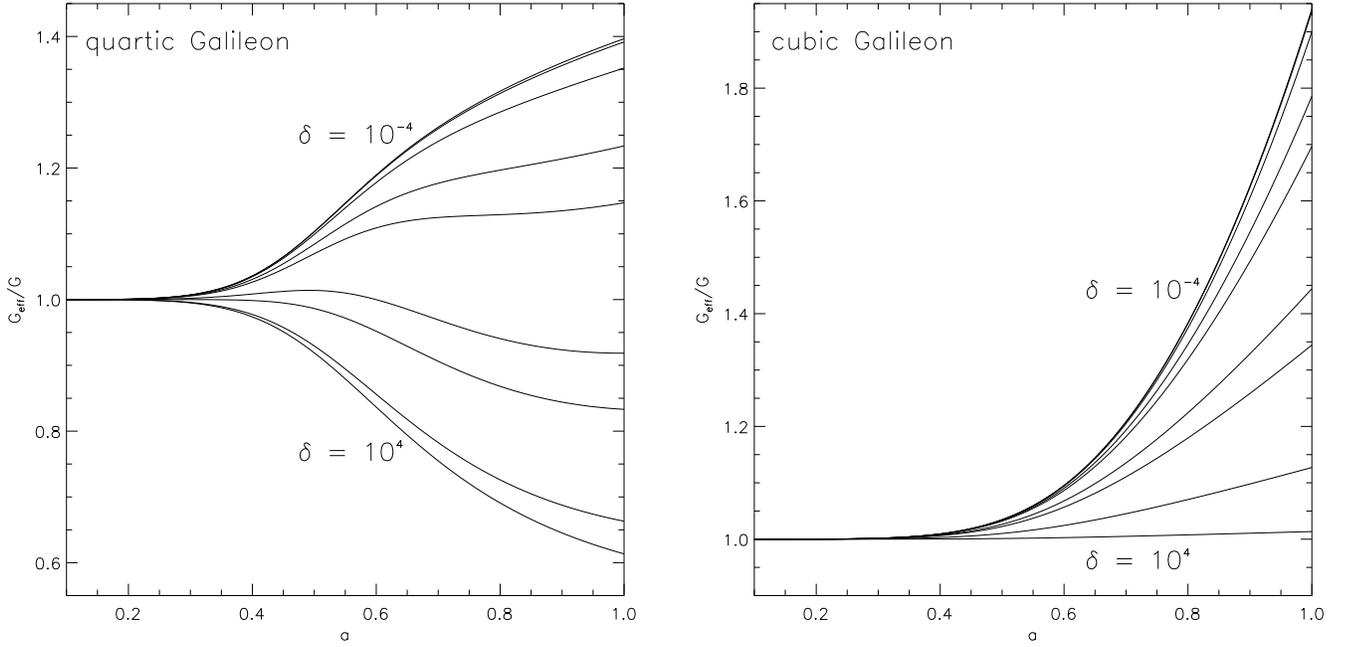}
\caption{The time evolution of $G_{\rm eff}/G$ for the quartic (left) and cubic (right) Galileon models, as predicted by a spherical calculation \cite{blbp2013}. This is shown for different overdensities, $\delta=10^{-4}$, $0.01$, $0.1$, $0.5$, $1.0$, $5.0$, $10.0$, $100.0$ and $10^4$ from top to bottom in each panel.}\label{fig:Geff}
\end{figure*}

As this is primarily a methodology paper, we will focus on some of the most important features of the quartic Galileon model. In particular, we shall look at the effect of the Galileon field on large-scale matter clustering, and try to understand it physically. More detailed studies, with higher-resolution simulations, will be left for future work.

\subsubsection{The matter power spectra}

One of the most useful statistics to quantify the clustering of matter is the two-point correlation function, or equivalently its Fourier transform -- the nonlinear matter power spectrum $P_{\delta\delta}(k)$ or $P(k)$. We used the publicly available {\sc powmes} \cite{powmes} code to measure $P(k)$ for all the simulations and the results are illustrated in Fig.~\ref{fig:pk}. To see more clearly the effects of modified gravity, we only show the relative difference of $P(k)$ in the full and linearised simulations from that in the corresponding QCDM simulation; all results are averaged over the five realisations and binned logarithmically in $k$.

The left panel of Fig.~\ref{fig:pk} shows the results from the $B=200$ $h^{-1}$Mpc simulation box. The red curves and circles, from top to bottom, show $P_{\rm linearised}(k)/P_{\rm QCDM}(k)-1$ at $a=1.0$, $0.8$ and $0.5$ respectively. From this we see that in the linearised simulations
\begin{itemize}
\item on linear scales ($k\lesssim0.1h$Mpc$^{-1}$) the simulation result agrees very well with the linear-theory prediction (black dashed horizontal lines) at all times, as expected;
\item on small scales mode coupling greatly enhances the clustering of matter compared to the linear-theory prediction, and this could be $\mathcal{O}(100\%)$ at $k\sim10h$Mpc$^{-1}$ today;
\item the deviation from QCDM increases with time, which is because the effective Newton constant $G_{\rm eff}$, which is given in Eq.~(\ref{eq:Geff}), grows with time, as we will see below.
\end{itemize}
The blue curves with squares, from bottom to top at $k\sim10$ $h$Mpc$^{-1}$, are results for the full quartic Galileon simulations at $a=1.0$, $0.8$ and $0.5$ respectively, and they follow very different behaviour from that of the linearised simulations. More explicitly,
\begin{itemize}
\item even though on large scales the full Galileon model predicts stronger clustering of matter than QCDM, the enhancement decreases as $k$ grows, and from $0.1\lesssim k\lesssim0.2h$Mpc$^{-1}$ on the matter clustering is actually {\it weaker} in the full model;
\item this trend grows with time;
\item at scales such as $k\sim0.03h$Mpc$^{-1}$, where linear perturbation theory is conventionally believed to still hold, the full Galileon simulations give a notably lower $P(k)$ than both linear theory and the linearised simulations at $a>0.5$, which indicates that nonlinearity is already important there; 
\item at $a\leq0.5$, however, the full Galileon simulations agree with linear-theory predictions very well on those large scales.
\end{itemize}

To confirm these observations, we did the same calculation for the $B=400h^{-1}$Mpc simulation box (not shown here to make the figure less busy), and found good agreement between the results from the two boxes.

To give physical explanations to the above observations, we need to know how the effective Newton constant $G_{\rm eff}$ changes with local matter density. This has been studied in \cite{blbp2013} in details for spherical top-hat configurations (see the colour-scale plot Fig.~3 of \cite{blbp2013}). Here we give a simplified account for the results and refer the interested readers to \cite{blbp2013} for more detail. Let us start with the definition of $G_{\rm eff}$:
\begin{itemize}
\item in the linearised simulations and linear perturbation theory, $G_{\rm eff}$ is given by Eq.~(\ref{eq:Geff}), and is independent of the matter density;
\item in the full Galileon model (spherical case), $G_{\rm eff}$ depends on local matter density inside the sphere, and is defined by
\begin{eqnarray}
G_{\rm eff}\delta\rho_m &=& G\left(\delta\rho_m+\rho_{\rm eff}\right),
\end{eqnarray}
where $\delta\rho_m$ is the matter density perturbation and $\rho_{\rm eff}$ is given by Eq.~(\ref{eq:rho_eff}).
\end{itemize}
The left panel of Fig.~\ref{fig:Geff} contains a few slices taken from Fig.~3 of \cite{blbp2013}, and it shows the time evolution of $G_{\rm eff}/G$ for a spherical tophat overdensity with different values of density, ranging from $\delta=10^{-4}$ (which is basically Eq.~(\ref{eq:Geff}), i.e, the linear-theory prediction) at the top to $\delta=10^4$ at the bottom. From this plot we can see that
\begin{itemize}
\item in the linearised simulations (the top curve) $G_{\rm eff}$ grows in time, regardless of the density value, which explains why matter clustering is stronger than in QCDM on all scales, especially small scales;
\item the situation becomes more complicated in the full simulations, in which $G_{\rm eff}$ sensitively depends on the density inside the sphere. As an example, even for $\delta=0.1$ there has been significant deviation of $G_{\rm eff}/G$ from its linear-theory value at $a\geq0.6$;
\item more interestingly, for $\delta\geq5.0$, $G_{\rm eff}/G$ is smaller than 1.0 at late times, implying a weakening of the total gravity. This is because, even though in high-density regions the spatial variations of the Galileon field have been efficiently smoothed out by the Vainshtein mechanism, so that the last two terms in Eq.~(\ref{eq:poisson_code_unit}) can be neglected, the time-dependent rescaling of $G$, caused by the curvature coupling of the Galileon field and specified by the coefficient $\alpha_1\alpha_4$ in Eq.~(\ref{eq:poisson_code_unit}), persists and makes $G_{\rm eff}<G$ at late times when $\alpha_1$ and $\alpha_4$ start to vary [cf.~Fig.~\ref{fig:background}]. 
\end{itemize}
As a comparison, we also plot the same results for the best-fit cubic Galileon model \cite{blhbp2013} (right panel of Fig.~\ref{fig:Geff}), in which $G_{\rm eff}/G$ is always larger than $1$ and approaches $1$ where matter density is high. Also, 
\begin{itemize}
\item the deviation of $G_{\rm eff}/G$ from unity starts at a later time than in the quartic case, and becomes much larger than the latter -- this indicates that density perturbations grow faster on large (linear) scales in the quartic model than in the cubic model;
\item at $\delta\geq0.1$ the difference in the values of $G_{\rm eff}/G$ given by linear theory and by the full cubic Galileon model is not as pronounced as in the quartic case -- this indicates that for the quartic Galileons the nonlinearty strikes in at lower densities and therefore affects larger scales more than in the cubic case (which explains why even at $k\sim0.02$ $h$Mpc$^{-1}$ the full model and linear-theory results do not fully agree with each other; c.f.~Fig.~\ref{fig:pk}).
\end{itemize}

The quartic Galileon model is different from all other modified gravity models we know of, in the sense that the deviations from standard gravity can be either positive or negative, depending on the local matter density or the length scale. This makes the model unique and therefore deserving of more detailed study. As this paper is mainly focused on the methodology, we will undertake such studies in a follow-up paper.

\subsubsection{The velocity divergence power spectra}

In the right panel of Fig.~\ref{fig:pk} we show the results for the velocity divergence power spectrum, which is defined by 
\begin{eqnarray}
P_{\theta\theta}(k) &\equiv& \langle|\mathbf{\theta_k}|^2\rangle,
\end{eqnarray}
where
\begin{eqnarray}
\mathbf{\theta_k} &\equiv& \frac{1}{(2\pi)^{3/2}}\int{\rm d}^3\mathbf{x}\theta(\mathbf{x})\exp(-i\mathbf{k}\cdot\mathbf{x}),
\end{eqnarray}
for the velocity divergence field $\theta(\mathbf{x})=\nabla\cdot\mathbf{v}/H$, and $\langle\cdots\rangle$ denotes the ensemble average.  Here we have assumed that the velocity filed is irrotational on the scales of interest, and can be described by the scalar $\theta(\mathbf{x})$ as a pure potential flow. This assumptions breaks down at small scales, however any potential effects of non-vanishing vorticity are beyond the scope of this work.

To measure the dark matter velocity field and its divergence from the particles' positions and velocities, we have used the Delaunay Tessellation Field Estimator (DTFE) code described in \cite{dtfe}. The use of the Delaunay tessellation has the advantage that the $\theta$ field calculated using this method is volume-averaged instead of mass-averaged. It also avoids the problem of empty cells containing no particles, which could appear in direct assignment methods to measure the velocity field \cite{ps2009}.

The results of $P_{\theta\theta}$ show qualitatively similar behaviour and patterns as we saw in the DGP simulations \cite{lzk2013} (Fig.~3 there). In particular, we can see that
\begin{itemize}
\item on scales as large as $k\sim0.04h$Mpc$^{-1}$, there is already substantial difference between results of the linearised and full simulations, which indicates that the nonlinearity due to the Galileon field can affect larger scales for the velocity field; indeed, this difference is much larger than what was found in the DGP simulations of \cite{lzk2013}, confirming that the model here has stronger nonlinearity;
\item at $k\sim0.2-1h$Mpc$^{-1}$, $P_{\theta\theta}$ is smaller in the full quartic Galileon model than in QCDM, because of the fact that $G_{\rm eff}/G<1$ in this regime;
\item for $k\geq1h$Mpc$^{-1}$, the deviation of $P_{\theta\theta}$ in the full Galileon model from the QCDM result decreases, and at $k\geq2h$ Mpc$^{-1}$ the former can even be slightly larger.
\end{itemize}
The above results paint a physical picture that in the linearised simulations the velocity field is boosted on all scales, while in the full simulations the situation becomes more complicated: on large scales the velocity divergence is enhanced but not by as much as the linear theory predicts because of the nonlinearity of the quartic Galileon model; on scales corresponding to the infall regions near halos and filaments, the weakened gravitational strength makes the velocity divergence smaller than in the case of standard gravity; on small scales corresponding to the substructures of halos, the virialisation process is modified by the weaker gravity: judging from the $P_{\delta\delta}$ results, one would expect that halos become less concentrated in the full simulations, because their potential is shallower but the particle velocities are not necessarily much lower than in QCDM -- this could lead to a more complicated behaviour of the velocity divergence on those scales.

The fact that the velocity field is seriously affected by the nonlinearity on (supposedly) linear scales ($k\sim0.04h$Mpc$^{-1}$) suggests that one should be careful in making conclusions regarding statistics such as the redshift space distortions by applying linear perturbation theory in models like the quartic Galileons. Fully consistent nonlinear simulations are essential to draw rigorous conclusions, and we will leave these to future work.

\section{Summary and conclusions}

\label{sect:summary}

\subsection{Summary of this work}

The quartic Galileon model is the next higher-order realisation of the Vainshtein mechanism to the cubic Galileon and DGP models, and the nonlinearity in its field equation is expected to restore GR near massive objects. Previous research has shown that this model could fit the combined CMB, supernovae and BAO data very well \cite{blbp2012,blsbp2013}, but studies have so far been restricted to the linear perturbation regime. This work is devoted to the development of the methodology needed to study the nonlinear structure formation in this model. To this end, we first derived the simplified Galileon field equation in the quasi-static and weak-field limits, which contains higher-order derivative-couplings of the Galileon field than in the cases of the DGP and cubic Galileon models.

We then generalised the algorithm for numerically solving the DGP and cubic Galileon equations to this model. The algorithm is based on the splitting of the second-order derivatives of the Galileon field $\varphi$ into a trace part ($\nabla^2\varphi$) and a traceless part ($\bar{\nabla}_i\bar{\nabla}_j\varphi$), so that the equation can be considered as a third-order algebraic equation of $\nabla^2\varphi$, the solutions to which can be obtained analytically. We explained which of the three branches of the solution is the physical one, and  discussed the required conditions for the physical solution to be real. The expression of $\nabla^2\varphi$ in terms of the density field and quantities containing $\bar{\nabla}_i\bar{\nabla}_j\varphi$ enabled us to make a very easy implementation of the Gauss-Seidel relaxation method to calculate $\varphi$ on (both regular and adaptively refined) meshes. Various possible numerical issues which are new to the quartic Galileon model were also discussed.

Our numerical Galileon-equation solver was then tested in different matter configurations, such as 1D sine and Gaussian density fields and spherical over(under)densities. We found in all these cases that the Galileon field can be accurately solved. We also identified a problem where no solution can be found for deep spherical underdensities, which is probably because of those terms which had been neglected under the quasi-static and weak-field limits. As the derivation of the full Galileon $N$-body equation is extremely complex and therefore beyond the scope of the current work, a simple and approximate cure of this problem was introduced to prevent the simulation from crashing. The same \tcr{fix} was introduced in the cubic Galileon simulations to avoid a similar problem, and was found to work reasonably well \cite{blhbp2013}.

To study how large-scale structures can be affected by a quartic Galileon field, we carried out a suite of $N$-body simulations, for both the full Galileon model and two of its variants under certain simplifying assumptions, with different box sizes and therefore mass (and force) resolutions. These simulations showed an interesting and unique feature of this model, namely it predicts stronger/weaker matter clustering than standard gravity (i.e., the QCDM model) on large/small scales. The behaviour would be completely different if one artificially suppresses the Vainshtein mechanism, such as in our linearised simulations, which show enhancement of the matter clustering on all scales (and especially at small scales).

We ascribed the distinct behaviour of the quartic Galileon model to the curvature coupling of the Galileon field, which makes the effective Newtonian constant, $G_{\rm eff}$, dependent on time. Similar to the DGP and cubic Galileon models, the Vainshtein mechanism could efficiently suppress the spatial variations of the Galileon field in high-density regions, but the Vainshtein mechanism itself does not affect the rescaling of $G_{\rm eff}$, and thus GR is {\it not} recovered in dense regions. This also has impacts on the large-scale velocity field, as illustrated by our results on the velocity divergence power spectrum.

\subsection{Discussions and outlook}

\subsubsection{{A summary of the Galileon models}}

The Galileon models (except for the cubic one) have been shown to fit the CMB, supernovae and BAO data rather well \cite{blsbp2013,blhbp2013,blbp2013}. However, linear theory predicts that these models generally strongly enhance the growth of structures on linear scales \cite{blsbp2013,blhbp2013,blbp2013}, which can be in tension with observations unless the galaxy bias is close to $1.0$ on those scales ($k\sim0.1$ $h$Mpc$^{-1}$). One then wonders if the nonlinearity can suppress this enhancement at $k\sim0.1h$Mpc$^{-1}$.

From the studies so far, we know the following:
\begin{itemize}
\item $N$-body simulations of \cite{blhbp2013} showed that for the best-fit cubic Galileon model, the nonlinearity does not affect scales larger than $k\sim0.1h$Mpc$^{-1}$; note that this model causes a large excess of the integrated Sachs Wolfe effect, which makes it disfavoured by CMB data \cite{blhbp2013};
\item $N$-body simulations in this work showed that nonlinearity can influence larger scales than it does in the cubic case -- as an example, at $k\sim0.02h$Mpc$^{-1}$ linear theory and simulations of the full model predict $\sim7\%$ and $\sim5\%$ enhancements of matter clustering respectively. But this still does not provide enough suppression of the linear matter power spectrum on large scales to make it comparable to the prediction of the best-fitting $\Lambda$CDM model (note however that the difference from $\Lambda$CDM is mainly due to the different matter density and expansion history than the Galileon models, rather than the law of gravitational interaction between particles \cite{blhbp2013,blbp2013}).
\item the study of spherical collapse in \cite{blbp2013} demonstrated that the quintic Galileon model admits no real physical solutions for spherical tophat overdensities whose density contrasts are higher than $\delta\sim\mathcal{O}(0.1-1)$ at late times.  As physical solutions do exist for the linearised quintic equation, it implies that nonlinearity has a more drastic effect on the quintic model than it does on the cubic and quartic models. 

We suspect that even if the approximations used to derive the quintic field equations are dropped (which is a highly challenging task itself) there is still no guarantee that this model has real physical solutions everywhere. As a result, we decide not to investigate this model further. 
\end{itemize}

Therefore, the nonlinearity becomes more important moving from the cubic to the quintic Galileons. Furthermore, one should bear in mind the problem in low-density regions identified for the cubic \cite{blhbp2013} and quartic (\cite{blbp2013} and this work) cases, which is another consequence of such nonlinearity. All these indicate that linear perturbation studies of the Galileon models can only be trusted for very large scales and/or very small density perturbations (e.g., the CMB), and that for most other cases fully consistent nonlinear investigations are essential.

{The time variation of $G_{\rm eff}$ in high-density regions is another important feature of the quartic (and quintic) Galileon model. This effect has been previously found in \cite{Galileon4} and in a slightly different context in \cite{bde2011}. If this conclusion also applies locally, such as in the Solar system, then the quartic Galileon model would be stringently constrained by local bounds on the time variation of $G$, $\dot{G}/G$ \cite{uzan2003}, such as those from the lunar laser ranging experiments \cite{wtb2004}. However, without a rigorous study, it is not clear how reliable our calculation, which is performed in a cosmological setup, can be used to make predictions about the behavior of the model in much smaller systems \cite{blbp2013}. It is clear that more work is needed to clarify this point\footnote{{Another point which needs more clarification, as pointed out in the original Galileon paper \cite{nrt2009}, is related to the sub-luminal propagation of the angular excitations of the Galileon field around the spherical solution, which makes the quasi-static approximation unreliable for certain Solar system observables. However, such restrictions for the model are obtained for spherical configurations and in the Minkowski spacetime, and therefore may not apply to the cosmological setup, in which sphericity is a poor approximation and the fully covariant Galileon equation \cite{dev2009} is significantly more complicated.}}.}



Admittedly, the above conclusions are made for the best-fit Galileon models using the CMB, supernovae and BAO data, and they can change if one adopts other Galileon and cosmological parameters. However, as the CMB and expansion data has been so powerful in constraining the Galileon parameters \cite{blsbp2013}, there seems to be no point to consider parameters which are drastically different from the best-fit values, and within the parameter space allowed by CMB data, we doubt that things would change fundamentally (for example, \cite{blbp2013} checked that the problem of no physical solutions for the quintic Galileon model persists for other reasonable parameter choices). 

\subsubsection{The numerical algorithm}

The numerical algorithm proposed in this work to solve the quartic Galileon equation is an extension of that for the DGP and cubic Galileon simulations in \cite{lzk2013,blhbp2013}, following the spirit of the operator-splitting technique \cite{cs2009}. By splitting $\nabla_i\nabla_j\varphi$ to $\nabla^2\varphi$ and $\bar{\nabla}_i\bar{\nabla}_j\varphi$, and solving the resulting third-order algebraic equation for $\nabla^2\varphi$, the equation can be reduced to the form $\nabla^2\varphi = \cdots$, the right-hand side of which contains complicated terms for the derivatives of $\varphi$ so that in principle these terms depend on $\varphi$. However, when the equation is discretised on a mesh, it turns out that the right-hand side no longer contains $\varphi_{i,j,k}$, which is the value of $\varphi$ in the central cell to be solved by the relaxation method. 

Therefore, as far as the relaxation method is concerned, the discrete version of $\nabla^2\varphi=\cdots$ is indeed a {\it linear} equation for $\varphi_{i,j,k}$, making the numerical implementation straightforward. Since the right-hand side of this equation (i.e., the `$\cdots$') does not contain $\varphi_{i,j,k}$, we do not even need a closed-form expression of it: all we need is its numerical values for all mesh cells. This implies that the algorithm here can be generalised to even higher-order theories, such as the {\it quintic} Galileon model, in which $\nabla^2\varphi$ satisfies a sixth-order algebraic equation \cite{blbp2013} and thus does not have analytical solution (this is different from the cubic and quartic cases). Of course, as the equation gets more nonlinear, we expect that the convergence properties will be worse in such higher-order theories, just as they are worse in the quartic than in the cubic models. 


\subsubsection{$N$-body implementations}

We have seen that the complexity of the field equations in the quartic Galileon model requires us to make certain approximations, but is is worth stressing that the approximations we made are not the only possibilities and one can follow alternatives as well.

An example is the treatment of $\bar{\nabla}_i\bar{\nabla}_j\Psi$ in the Galileon field equation: although there are different ways to do this, none of them is ideal. The approach we followed is a practical one, in which we took the values of $\Psi$ from the previous time step as a source term for the Galileon equation at the current step. We can think of other variants of this approach, e.g., the following steps 
\begin{enumerate}
\item split $\bar{\nabla}_i\bar{\nabla}_j\Psi$ into $\bar{\nabla}_i\bar{\nabla}_j\Psi_{\rm GR}$ and the terms involving derivatives of $\varphi$, as in Eq.~(\ref{eq:ij_traceless_tracker2}),
\item put this splitting into the Galileon equation, which now contains $\bar{\nabla}_i\bar{\nabla}_j\Psi_{\rm GR}$ (and no longer $\bar{\nabla}_i\bar{\nabla}_j\Psi$), as well as a few more (higher-order) $\varphi$-derivative terms,
\item solve $\Psi_{\rm GR}$ before solving the Galileon equation so that when the latter is being solved the value of $\bar{\nabla}_i\bar{\nabla}_j\Psi_{\rm GR}$ at the {\it current} step is already known, and
\item use the previous-step values of the $\varphi$-derivative terms in Eq.~(\ref{eq:ij_traceless_tracker2}) in the Galileon equation (which then does {\it not} become higher-order).
\end{enumerate}
As not all the $\varphi$-derivative terms in Eq.~(\ref{eq:ij_traceless_tracker2}) are higher-order, one can even decide to absorb some of them, e.g. $\nabla^2\varphi\bar{\nabla}_i\bar{\nabla}_j\varphi$, into a redefinition of the $\beta$-coefficients in the Galileon equation.

Note that such a modified implementation will likely also change how the total gravitational force is computed (as $\Psi$ is not being computed it this), and one then needs to use Eq.~(\ref{eq:force}) to calculate the gravitational force. {We tested this and found that the result only slightly differs from that in Fig.~\ref{fig:pk} for $0.2\lesssim k/(h{\rm Mpc}^{-1})\lesssim4$ (the difference is significantly smaller than the 1$\sigma$ error bars therein). As the method described in the main text requires us to store fewer quantities, it is more memory efficient and therefore better for large simulations.}

\subsubsection{Future work}

This paper is devoted to the development of methodology, and therefore does not contain an in-depth analysis of the effect the quartic Galileon field may have on different cosmological observables. Because a distinct feature of this model is that gravity is strengthened on large scales (or in low-density regions) and weakened on small scales (or in dense regions), we expect that it could have a big impact on the formation and properties of dark matter halos (such as the halo concentrations). On the other hand, we also show that the velocity field is more strongly affected by the Vainshtein mechanism, which can have important effects on the redshift space distortions. It would thus be of interests to run high-resolution or large-box simulations for this model in the future, and study these issues in greater details.

Moreover, the development of the excursion set approach \cite{bcek1991,blbp2013} provides a powerful complementary tool to analyse nonlinear structure formation in the quartic Galileon models. Its predictions for the halo mass function and halo bias etc. can be tested with high-resolution simulations. With some extra work, one can also build up a halo model \cite{cs2002} for the quartic Galileons, which will enable us to predict other observables, such as the matter power spectrum, and compare with the full simulations. 

\begin{acknowledgments}

We thank Romain Teyssier for helpful discussions on the details of the {\sc ramses} code. BL is supported by the Royal Astronomical Society and Durham University. AB is supported by the FCT-Portugal via grant SFRH/BD/75791/2011. WAH is supported by the Polish National Science Centre through grant DEC-2011/01/D/ST9/01960 and the European Research Council (ERC) Advanced Investigator grant of C.~S.~Frenk, {\sc cosmiway}. KK is supported by STFC grant ST/H002774/1, the Leverhulme Trust and the ERC. GBZ is supported by the Dennis Sciama Fellowship from the University of Portsmouth. This work has also been partially supported by the European Union FP7 ITN INVIS- IBLES (Marie Curie Actions, PITN- GA-2011- 289442) and STFC. The simulations described in this paper were carried out on the {\sc cosma4} supercomputer at the Institute for Computational Cosmology, Durham University. 

\end{acknowledgments}

\begin{appendix}

\section{Useful expressions}

\label{appendix:a}

In this appendix we give the expressions of some intermediate quantities used to derive the Einstein and Galileon equations in the quasi-static limit. These are for future references and for the interested readers to cross check.

As mentioned in the main text, in this paper we work with the Newtonian-gauge metric of Eq.~(\ref{eq:metric}). Under the quasi-static and weak-field approximations, the only nonzero components of the Christoffel symbol are\footnote{Note that terms such as $\dot{\Psi}$ are neglected in these approximations.}
\begin{eqnarray}
\Gamma^0_{0i} &\doteq& \Psi_{,i},\nonumber\\
\Gamma^0_{ij} &\doteq& a^2H\gamma_{ij},\nonumber\\
\Gamma^i_{00} &\doteq& \frac{1}{a^2}\gamma^{ij}\Psi_{,j},\nonumber\\
\Gamma^i_{0j} &\doteq& H\delta^i_j,\nonumber\\
\Gamma^i_{jk} &\doteq& -\Phi_{,k}\delta^i_j - \Phi_{,j}\delta^i_k + \gamma^{il}\gamma_{jk}\Phi_{,l},
\end{eqnarray}
in which $_{,i}\equiv\partial/\partial x^i$ ($x^i$ being the comoving coordinate) and $\delta^i_j$ is the Kronecker delta. 

With these quantities, the components of curvature tensors which are used in the derivation of the field equations are
\begin{eqnarray}
R_{00} &\doteq& \frac{1}{a^2}\Psi^{,i}_{\ ,i} - 3\left(\dot{H}+H^2\right),\nonumber\\
R_{ij} &\doteq& \left(\dot{H}+3H^2\right)a^2\gamma_{ij} + \Phi^{,k}_{\ ,k}\gamma_{ij} + \left(\Phi-\Psi\right)_{,ij},\nonumber\\
R &\doteq& \frac{2}{a^2}\left(\Psi-2\Phi\right)^{,i}_{\ ,i} - 6\left(\dot{H}+2H^2\right),\nonumber\\
G_{00} &\doteq& \frac{2}{a^2}\Phi^{,i}_{\ ,i} + 3H^2,\nonumber\\
G_{ij} &\doteq& \gamma_{ij}\left(\Psi-\Phi\right)^i_{\ ,i} + \left(\Phi-\Psi\right)_{,ij},
\end{eqnarray}
and the elementary derivative terms of the Galileon field that appear in the field equations are
\begin{eqnarray}
\Box\varphi &\doteq& \ddot{\varphi}+3H\dot{\varphi}-\frac{1}{a^2}\varphi^{,i}_{\ ,i},\nonumber\\
\nabla^\mu\varphi\nabla_\mu\varphi &\doteq& \dot{\varphi}^2,\nonumber\\
\nabla^\mu\varphi\nabla^\nu\varphi\nabla_\mu\nabla_\nu\varphi &\doteq& \dot{\varphi}^2\ddot{\varphi},\nonumber\\
\nabla^\mu\nabla^\nu\varphi\nabla_\mu\nabla_\nu\varphi &\doteq& \ddot{\varphi}^2 + 3H^2\dot{\varphi}^2 - \frac{2}{a^2}H\dot{\varphi}\varphi^{,i}_{\ ,i}\nonumber\\
&& + \frac{1}{a^4}\varphi^{,ij}\varphi_{,ij},
\end{eqnarray}
and
\begin{eqnarray}
&&\nabla^\mu\nabla^\nu\varphi\nabla_\nu\nabla_\rho\varphi\nabla^\rho\nabla_\mu\varphi\nonumber\\
&\doteq& \ddot{\varphi}^3 + 3H^3\dot{\varphi}^3 - \frac{3}{a^2}H^2\dot{\varphi}^2\varphi^{,i}_{\ ,i} + \frac{3}{a^4}H\dot{\varphi}\varphi^{,ij}\varphi_{,ij}\nonumber\\
&&  - \frac{1}{a^6}\varphi^{,ij}\varphi_{,jk}\varphi^{,k}_{\ ,i}.
\end{eqnarray}

\section{Discretisation of derivative terms}

\label{sect:derivative_discretisation}

In this appendix we show the discrete versions of the derivative coupling terms involved in the Galileon field equations. These are (note again that the subscripts $_{i,j,k}$ are the cell indices along the $x,y,z$ directions)

\begin{widetext}
\begin{eqnarray}
\bar{\nabla}^i\bar{\nabla}^j\varphi\bar{\nabla}_i\bar{\nabla}_j\varphi 
&=&  \frac{1}{3h^4}\left(\varphi_{i+1,j,k}+\varphi_{i-1,j,k}\right)\left(2\varphi_{i+1,j,k}+2\varphi_{i-1,j,k}-\varphi_{i,j+1,k}-\varphi_{i,j-1,k}-\varphi_{i,j,k+1}-\varphi_{i,j,k-1}\right)\nonumber\\
&& + \frac{1}{3h^4}\left(\varphi_{i,j+1,k}+\varphi_{i,j-1,k}\right)\left(2\varphi_{i,j+1,k}+2\varphi_{i,j-1,k}-\varphi_{i+1,j,k}-\varphi_{i-1,j,k}-\varphi_{i,j,k+1}-\varphi_{i,j,k-1}\right)\nonumber\\
&& + \frac{1}{3h^4}\left(\varphi_{i,j,k+1}+\varphi_{i,j,k-1}\right)\left(2\varphi_{i,j,k+1}+2\varphi_{i,j,k-1}-\varphi_{i+1,j,k}-\varphi_{i-1,j,k}-\varphi_{i,j+1,k}-\varphi_{i,j-1,k}\right)\nonumber\\
&& + \frac{1}{8h^4}\left(\varphi_{i+1,j+1,k}+\varphi_{i-1,j-1,k}-\varphi_{i+1,j-1,k}-\varphi_{i-1,j+1,k}\right)^2\nonumber\\
&& + \frac{1}{8h^4}\left(\varphi_{i+1,j,k+1}+\varphi_{i-1,j,k-1}-\varphi_{i+1,j,k-1}-\varphi_{i-1,j,k+1}\right)^2\nonumber\\
&& + \frac{1}{8h^4}\left(\varphi_{i,j+1,k+1}+\varphi_{i,j-1,k-1}-\varphi_{i,j+1,k-1}-\varphi_{i,j-1,k+1}\right)^2,
\end{eqnarray}
\begin{eqnarray}
&&\bar{\nabla}^i\bar{\nabla}^j\varphi\bar{\nabla}_i\bar{\nabla}_j\Psi\nonumber\\
&=&  \frac{1}{8h^4}\left(\varphi_{i+1,j+1,k}+\varphi_{i-1,j-1,k}-\varphi_{i+1,j-1,k}-\varphi_{i-1,j+1,k}\right)\left(\Psi_{i+1,j+1,k}+\Psi_{i-1,j-1,k}-\Psi_{i+1,j-1,k}-\Psi_{i-1,j+1,k}\right)\nonumber\\
&& + \frac{1}{8h^4}\left(\varphi_{i+1,j,k+1}+\varphi_{i-1,j,k-1}-\varphi_{i+1,j,k-1}-\varphi_{i-1,j,k+1}\right)\left(\Psi_{i+1,j,k+1}+\Psi_{i-1,j,k-1}-\Psi_{i+1,j,k-1}-\Psi_{i-1,j,k+1}\right)\nonumber\\
&& + \frac{1}{8h^4}\left(\varphi_{i,j+1,k+1}+\varphi_{i,j-1,k-1}-\varphi_{i,j+1,k-1}-\varphi_{i,j-1,k+1}\right)\left(\Psi_{i,j+1,k+1}+\Psi_{i,j-1,k-1}-\Psi_{i,j+1,k-1}-\Psi_{i,j-1,k+1}\right)\nonumber\\
&& + \frac{1}{3h^4}\left(\Psi_{i+1,j,k}+\Psi_{i-1,j,k}\right)\left(2\varphi_{i+1,j,k}+2\varphi_{i-1,j,k}-\varphi_{i,j+1,k}-\varphi_{i,j-1,k}-\varphi_{i,j,k+1}-\varphi_{i,j,k-1}\right)\nonumber\\
&& + \frac{1}{3h^4}\left(\Psi_{i,j+1,k}+\Psi_{i,j-1,k}\right)\left(2\varphi_{i,j+1,k}+2\varphi_{i,j-1,k}-\varphi_{i+1,j,k}-\varphi_{i-1,j,k}-\varphi_{i,j,k+1}-\varphi_{i,j,k-1}\right)\nonumber\\
&& + \frac{1}{3h^4}\left(\Psi_{i,j,k+1}+\Psi_{i,j,k-1}\right)\left(2\varphi_{i,j,k+1}+2\varphi_{i,j,k-1}-\varphi_{i+1,j,k}-\varphi_{i-1,j,k}-\varphi_{i,j+1,k}-\varphi_{i,j-1,k}\right),
\end{eqnarray}
\begin{eqnarray}
&&\bar{\nabla}_i\bar{\nabla}_j\varphi\bar{\nabla}^j\bar{\nabla}^k\varphi\bar{\nabla}_k\bar{\nabla}^i\varphi\nonumber\\
&=&  \frac{1}{9h^6}\left(\varphi_{i+1,j,k}+\varphi_{i-1,j,k}\right)\left[2\left(\varphi_{i+1,j,k}+\varphi_{i-1,j,k}\right)^2+\left(\varphi_{i,j+1,k}+\varphi_{i,j-1,k}\right)^2+\left(\varphi_{i,j,k+1}+\varphi_{i,j,k-1}\right)^2\right]\nonumber\\
&& + \frac{1}{9h^6}\left(\varphi_{i,j+1,k}+\varphi_{i,j-1,k}\right)\left[2\left(\varphi_{i,j+1,k}+\varphi_{i,j-1,k}\right)^2+\left(\varphi_{i+1,j,k}+\varphi_{i-1,j,k}\right)^2+\left(\varphi_{i,j,k+1}+\varphi_{i,j,k-1}\right)^2\right]\nonumber\\
&& + \frac{1}{9h^6}\left(\varphi_{i,j,k+1}+\varphi_{i,j,k-1}\right)\left[2\left(\varphi_{i,j,k+1}+\varphi_{i,j,k-1}\right)^2+\left(\varphi_{i+1,j,k}+\varphi_{i-1,j,k}\right)^2+\left(\varphi_{i,j+1,k}+\varphi_{i,j-1,k}\right)^2\right]\nonumber\\
&& -  \frac{2}{9h^6}\left(\varphi_{i+1,j,k}+\varphi_{i-1,j,k}\right)^2\left(\varphi_{i,j+1,k}+\varphi_{i,j-1,k}+\varphi_{i,j,k+1}+\varphi_{i,j,k-1}\right)\nonumber\\
&& -  \frac{2}{9h^6}\left(\varphi_{i,j+1,k}+\varphi_{i,j-1,k}\right)^2\left(\varphi_{i+1,j,k}+\varphi_{i-1,j,k}+\varphi_{i,j,k+1}+\varphi_{i,j,k-1}\right)\nonumber\\
&& -  \frac{2}{9h^6}\left(\varphi_{i,j,k+1}+\varphi_{i,j,k-1}\right)^2\left(\varphi_{i+1,j,k}+\varphi_{i-1,j,k}+\varphi_{i,j+1,k}+\varphi_{i,j-1,k}\right)\nonumber\\
&& -  \frac{2}{9h^6}\left(\varphi_{i+1,j,k}+\varphi_{i-1,j,k}\right)\left(\varphi_{i,j+1,k}+\varphi_{i,j-1,k}-\varphi_{i,j,k+1}-\varphi_{i,j,k-1}\right)^2\nonumber\\
&& -  \frac{2}{9h^6}\left(\varphi_{i,j+1,k}+\varphi_{i,j-1,k}\right)\left(\varphi_{i+1,j,k}+\varphi_{i-1,j,k}-\varphi_{i,j,k+1}-\varphi_{i,j,k-1}\right)^2\nonumber\\
&& -  \frac{2}{9h^6}\left(\varphi_{i,j,k+1}+\varphi_{i,j,k-1}\right)\left(\varphi_{i+1,j,k}+\varphi_{i-1,j,k}-\varphi_{i,j+1,k}-\varphi_{i,j-1,k}\right)^2\nonumber\\
&& + \frac{1}{16h^6}\left(\varphi_{i+1,j,k}+\varphi_{i-1,j,k}+\varphi_{i,j+1,k}+\varphi_{i,j-1,k}-2\varphi_{i,j,k+1}-2\varphi_{i,j,k-1}\right)\nonumber\\
&&~~~~~~~~~~\times\left(\varphi_{i+1,j+1,k}+\varphi_{i-1,j-1,k}-\varphi_{i+1,j-1,k}-\varphi_{i-1,j+1,k}\right)^2\nonumber\\
&& + \frac{1}{16h^6}\left(\varphi_{i+1,j,k}+\varphi_{i-1,j,k}+\varphi_{i,j,k+1}+\varphi_{i,j,k-1}-2\varphi_{i,j+1,k}-2\varphi_{i,j-1,k}\right)\nonumber\\
&&~~~~~~~~~~\times\left(\varphi_{i+1,j,k+1}+\varphi_{i-1,j,k-1}-\varphi_{i+1,j,k-1}-\varphi_{i-1,j,k+1}\right)^2\nonumber\\
&& + \frac{1}{16h^6}\left(\varphi_{i,j+1,k}+\varphi_{i,j-1,k}+\varphi_{i,j,k+1}+\varphi_{i,j,k-1}-2\varphi_{i+1,j,k}-2\varphi_{i-1,j,k}\right)\nonumber\\
&&~~~~~~~~~~\times\left(\varphi_{i,j+1,k+1}+\varphi_{i,j-1,k-1}-\varphi_{i,j+1,k-1}-\varphi_{i,j-1,k+1}\right)^2\nonumber\\
&& + \frac{3}{32h^6}\left(\varphi_{i+1,j+1,k}+\varphi_{i-1,j-1,k}-\varphi_{i+1,j-1,k}-\varphi_{i-1,j+1,k}\right)\left(\varphi_{i+1,j,k+1}+\varphi_{i-1,j,k-1}-\varphi_{i+1,j,k-1}-\varphi_{i-1,j,k+1}\right)\nonumber\\
&&~~~~~~~~~~\times\left(\varphi_{i,j+1,k+1}+\varphi_{i,j-1,k-1}-\varphi_{i,j+1,k-1}-\varphi_{i,j-1,k+1}\right).\ \ \
\end{eqnarray}
\end{widetext}

We can observe two important features of these terms:
\begin{itemize}
\item none of them contain $\varphi_{i,j,k}$;
\item all of them vanish in the homogeneous case, where $\varphi$ is the same everywhere.
\end{itemize}
The first property, in particular, is the benefit of defining the barred derivative $\bar{\nabla}_i\bar{\nabla}_j\varphi$.

\end{appendix}

\end{document}